\documentclass{iopjournal}

\usepackage{subcaption}
\usepackage{amsmath} 
\usepackage{ragged2e}

\begin{document}

\articletype{Paper} 

\title{Stochastic single-stage stellarator optimization using fixed-boundary equilibria}

\author{Pedro F. Gil$^{1,*}$, Jason Smoniewski$^{2,3}$, Rogerio Jorge$^{4,5}$, Paul Huslage$^1$ and Eve V. Stenson$^1$}

\affil{$^1$Max-Planck-Institut für Plasmaphysik, 85748 Garching, Germany}

\affil{$^2$Max-Planck Institute for Plasma Physics,
 17491 Greifswald, Germany}
 
\affil{$^3$Plasma Science and Fusion Center, Massachusetts Institute of Technology, Cambridge, Massachusetts 02139, USA}

\affil{$^4$Department of Physics, University of Wisconsin-Madison, Madison, Wisconsin 53706, USA}

\affil{$^5$Instituto de Plasmas e Fusão Nuclear, Instituto Superior Técnico, Universidade de Lisboa, 1049-001 Lisboa, Portugal}

\affil{$^*$Author to whom any correspondence should be addressed.}

\email{pedro.gil@ipp.mpg.de}

\begin{abstract}
\justifying
\noindent In this paper, single-stage stellarator optimization is combined with stochastic coil optimization to improve the robustness of the stellarator as compared to deterministic methods. The plasma boundary, solved with an MHD solver in fixed-boundary mode, is linked to a set of randomly perturbed coils via the squared flux. The optimizer avoids sharp local minima and can reach improved configurations. Two different configurations obtained with our method, one quasi-axisymmetric and one quasi-helically symmetric, are compared against both the standard stochastic stage II method and the single-stage method. The new configurations shown here yield improved squared flux, quasisymmetry, and particle loss following a posteriori perturbation of the coils. 
\end{abstract}

\section{Introduction}
\justifying
In recent years, interest in stellarators as a viable option for achieving commercial fusion has increased. Stellarators are toroidal devices that confine a plasma using 3D-shaped coils. These coils generate a magnetic field with a rotational transform without significant current in the plasma and allow for steady-state operation. Due to relaxing the requirement for axisymmetry, stellarator optimization has broad potential to meet multiple physics requirements, such as minimizing neoclassical transport, mitigating MHD instabilities, and improving the confinement of fast particles \cite{w7x_beidler, goodman_prx, LandremanPaul}. 

One main disadvantage of stellarators is their complexity to build. This also means that it is difficult to meet building tolerances. A notable example is the National Compact Stellarator eXperiment (NCSX) in the USA, which was canceled primarily because of difficulty meeting the tolerances \cite{ncsx}. Tolerances here refer to the allowed manufacturing and assembly geometrical deviations of the coils. Moreover, although the construction of Wendelstein 7-X has proven to be a success, tolerances were relatively small compared to the size of the coils: 2 mm allowed deviations for coils with 3.5 m diameter. These were determined to keep the relative magnetic field deviations within $10^{-4}$ \cite{Pedersen2016, andreeva_2004}. These results have sparked extensive work on ways to improve the shape of the non-planar coils and design plasma shapes that simultaneously meet physics requirements and relax engineering constraints \cite{Jorge_2023, Jorge_2024, Giuliani_2024, Giuliani_2024_quasr_exploration, Kaptanoglu_2025, LandremanPaul, Landreman_2018_shape_gradient, Paul_Antonsen_Landreman_Cooper_2020_shape_gradient}. 

Recent work has shown that it is possible to reduce neoclassical transport in stellarators across the volume and achieve values similar to those of axisymmetric devices by targeting quasisymmetry, which acts as a proxy for neoclassical transport \cite{LandremanPaul}. Quasisymmetry is a property of the magnetic field amplitude that, if invariant along a given Boozer coordinate, allows for the confinement of collisionless trapped particles. This method was first used in a stage I approach, where solely the plasma equilibrium is optimized. Typically, this is followed by a stage II optimization, where a magnetostatic inverse problem is solved to determine coils that produce a field matching the previously-found equilibrium surface. The main disadvantage of this sequential two-stage approach is that the equilibrium optimization is done 'blindly'. It often results in plasmas that are far too complex to find buildable coils that accurately reproduce the plasma shape. Moreover, the inverse problem of finding coils is ill-posed, as the number of coil degrees of freedom is far larger than the number of equations that are available to regularize the coils. The consequence is that for one equilibrium, countless different coil sets can be found. This makes the entire optimization procedure complex. 

Two notable developments in this domain are single-stage methods \cite{henneberg_single_stage,giuliani_single_stage, Jorge_2023} and stochastic coil optimization \cite{Wechsung_2, Wechsung_2022, Lobsien_2018}. The former combines the shaping of the plasma while simultaneously modifying the coils. This is needed since stage II, although effective at meeting the coil engineering constraints, such as maximum curvature, length, or torsion, it does so in detriment of field accuracy, due to poor compatibility between plasma and coils. 
Additionally, stochastic optimization has been developed to address the challenges associated with engineering tolerances. The optimization targets a cloud of perturbed coils surrounding the central coil set, rather than a single coil. For a ``narrow'' minimum, the cloud of perturbed coils will increase the value of the optimization cost function, as compared to a ``broader'' minimum. This means that, under coil perturbations, the value of the cost function should not increase as much as for configurations optimized with traditional methods, resulting in a more robust stellarator. The current paper combines the advantages of both these methods. 

This article is organized as follows: In Section \ref{sec:optimization_methods}, the individual methods of single-stage and stochastic coil optimization are briefly explained, and then the merging of the two is presented. In Section \ref{sec:numerical_results}, the stochastic single-stage optimized configurations are presented and compared to the deterministic version. Moreover, two stellarator configurations, one quasi-axisymmetric (QA) and one quasi-helically symmetric (QH), were optimized and checked against quasisymmetry, coil accuracy, robustness, and alpha particle confinement metrics. We conclude with Section \ref{sec:conclusion}.

\section{Optimization Methods}
\label{sec:optimization_methods}
\subsection{\textit{Single Stage with Fixed Boundary Equilibrium}}
In the single-stage approach to stellarator optimization, the shape of both plasma and coils are optimized simultaneously. The method adopted in this paper was developed by R. ~Jorge et al.~\cite{Jorge_2023} with the optimization software \texttt{SIMSOPT} \cite{landreman_simsopt_2021}, where the plasma is represented via a magnetohydrodynamic (MHD) equilibrium from the Variational Moments Equilibrium Code (VMEC)\cite{hirshman_steepestdescent_1983-1}. VMEC solves for ideal MHD equilibria by minimizing the MHD potential energy $W$ on a series of toroidally nested flux surfaces via a steepest descent method, where 

\begin{equation}
    W = \int \left(\frac{p}{\gamma -1} + \frac{\vert \Vec{B} \vert^2 }{2\mu_0} \right)dV .
\label{eq:VMEC}
\end{equation}

Here $p$ is the plasma pressure, $B$ the magnetic field, $\gamma$ the adiabatic index, and V the volume considered.

For a free-boundary problem there is no prescribed boundary, instead the coils and their currents are given, then an iterative process is used to find the correspondent plasma boundary. While it is more physically accurate, a free-boundary problem is more computationally expensive and more difficult to converge. For a fixed-boundary problem, the computational domain is given by the LCFS, provided as an input. This work uses VMEC to solve fixed-boundary problems, where the surface is parameterized in cylindrical coordinates via:

\begin{subequations}
\begin{eqnarray}
    R(\theta, \phi) = \sum_{m=0}^{M_{pol}}\sum_{n=-N_{tor}}^{N_{tor}} \left[ a_{m,n}\cos(m\theta-n_{fp}n\phi) + b_{m,n}\sin(m\theta-n_{fp}n\phi) \right] ,\\
    Z(\theta, \phi) = \sum_{m=0}^{M_{pol}}\sum_{n=-N_{tor}}^{N_{tor}} \left[ c_{m,n}\cos(m\theta-n_{fp}n\phi) + d_{m,n}\sin(m\theta-n_{fp}n\phi) \right] .
\end{eqnarray}
 \label{eq:surface}
\end{subequations}

Here, $\theta$ is the poloidal angle, $\phi$ is the toroidal angle, and
($M_{pol}$,$N_{tor}$) are the number of poloidal and toroidal modes. The cosine and sine coefficients $x_{eq}$=($a_{m,n}, b_{m,n}, c_{m,n}, d_{m,n}$) define the shape of the surface and are the degrees of freedom  for optimization.



Quasisymmetry is highly desirable and is added in this work as an objective function. This is done through the two-term metric for quasisymmetry \cite{Rodríguez_Paul_Bhattacharjee_2022}:

\begin{equation}
    f_{QS} = \sum_{s_i} w_i\biggr \langle \left[ \frac{1}{|\Vec{B}|^3}\left( (N-\iota M) \Vec{B} \times \Vec{\nabla} B \cdot \Vec{\nabla} \psi - (MG+NI) \Vec{B} \cdot \Vec{\nabla} B \right) \right]^2\biggr \rangle .
    \label{eq:quasisymmetry}
\end{equation}

Where $\psi$ is the radial flux coordinate, G and I are the poloidal and toroidal currents respectively, and $s_i$ correspond to the surfaces where the metric is evaluated. On top of the quasisymmetry metric, other targets were used in the optimization, namely aspect ratio $A$ and rotational transform $\iota$:

\begin{subequations}
    \begin{gather}
        f_{\iota} = (\iota - \iota_{T})^2 ,\\
        f_{A} = (A - A_{T})^2 .
    \end{gather}
    \label{eq:equilibrium_functions}
\end{subequations}

Where $A_{T}$ and $\iota_{T}$ are the target values set by the user in the optimization.  

The coils are modeled as 1D current-carrying filaments, defined by a Fourier decomposition of the XYZ components:

\begin{equation}
x^i(\theta)=\sum_{m=0}^{M_{coil}}x_{c,m}^icos(m\theta)+\sum_{n=0}^{M_{coil}}x_{s,n}^isin(n\theta) ,
\label{eq:coils_fourier}
\end{equation}

where, $x^i$ is one of the XYZ components, $M_{coil}$ sets the number of Fourier modes discretizing the coil. The coefficients $x_{coil}$=($x^i_{c,m}, x^i_{s,n}$) are the degrees of freedom that are modified in the optimization to shape the coils. The shaping is driven by minimizing the square of the magnetic flux through the equilibrium surface, which is traditionally a measure of field error referred to as the squared flux, however here it can be interpreted as a coupling term between the coils and the plasma surface:

\begin{equation}
    f_{SF} = \frac{1}{2}\int_{S} \frac{\vert \Vec{B} \cdot \Vec{n}\vert^2}{\vert \Vec{B} \vert ^2}dS .
    \label{eq:squared_flux}
\end{equation}

$\Vec{B}$ is calculated on the surface's quadrature points through the Biot-Savart law, $\Vec{n}$ is the surface normal, and $S$ represents, the LCFS. Solely minimizing the squared flux is insufficient to obtain buildable coils, as it is an ill-posed problem \cite{stellarator_intro}. Therefore, further regularization functions are defined as:

\begin{subequations}
	\begin{gather}
    f_{l} = \frac{1}{2}\left(\max\left(\sum_{i=1}^{N} L_i - L_{0},0\right)\right)^2 \label{eq:length_penalty},\\
    f_\text{msc} = \frac{1}{2}\left(\max \left(\sum_{i=1}^{N}\frac{1}{L_i}\int_{\Gamma_i} \kappa^2 dl - K_0,0\right)\right)^2, \\
    f_\text{curv} = \sum_{i=1}^{N}\frac{1}{2}\int_{\Gamma_i} \max(\kappa - \kappa_0,0)^2 dl, \\
    f_{cs} = \sum_{i = 1}^{N}\int_{\Gamma_i} \int_{S} \max(0, d_{0}^{cs} - \| \mathbf{r}_i - \mathbf{s} \|_2)^2 ~dl_i ~dS,\\
    f_{cc} = \sum_{i = 1}^{N} \sum_{j = 1}^{i-1} \int_{\Gamma_i} \int_{\Gamma_j} \max(0, d_{0}^{cc} - \| \mathbf{r}_i - \mathbf{r}_j \|_2)^2 ~dl_j ~dl_i, \\
	f_\text{link}(\Gamma_i, \Gamma_j) = \frac{1}{4\pi} \left| \oint_{\Gamma_i}\oint_{\Gamma_j}\frac{\textbf{r}_i - \textbf{r}_j}{|\textbf{r}_i - \textbf{r}_j|^3} (d\textbf{r}_i \times d\textbf{r}_j) \right|.
        \end{gather}
\label{eq:coil_functions}
\end{subequations}

Here, N is the total number of coils, $\Gamma_i$ is the shape or path of coil $i$, and $i,j$ iterate over different coils.
$f_{l}$ penalizes the length of all the coils in one half-field period and $L_0$ is the maximum length threshold. $f_{msc}$ the mean square curvature where $\kappa$ is the local curvature and $K_0$ the corresponding threshold.  $f_{curv}$ targets the maximum curvature of the coils where $\kappa_0$ is a threshold set by the user. $f_{cs}$ optimizes for the coil to surface distance where $d_{0}^{cs}$ is the minimal accepted distance also set by the user, $\textbf{r}_i$ and $s$ are points on coil $i$ and surface, and similarly for $f_{cc}$ and $d_{0}^{cc}$. $f_{link}$ calculates the Gauss linking number of two curves $i$ and $j$, when it is zero it means that no curves are interlinked. 

Single-stage optimization, by definition, encompasses both the equilibrium targets $J_{eq}$ and coil targets $J_{coil}$ in one single minimizing function:

\begin{subequations}
    \begin{gather}
    J = J_{eq} + J_{coil} ,\\
    J_{eq} = w_{QS}f_{QS}+w_{\iota}f_{\iota} + w_Af_{A} ,\\
    J_{coil} = w_{SF}f_{SF} + w_{coil} \left( \sum_i w_i f_i \right) ,
    \end{gather}
    \label{eq:single_stage}
\end{subequations}

where $w_i$ are user-set hyperparameters. Note that $J_{eq}$ is a weighted sum of all the objective functions of equations \ref{eq:equilibrium_functions}, and $J_{coil}$ also adds similarly the functions in equations \ref{eq:coil_functions} as well as the squared flux function $J_{SF}$. This method of assembling both objects under one function couples the coils to the surface via the squared flux. $w_{coil}$ is a way to regulate this coupling.

\subsection{\textit{Stochastic Optimization}}

In this section, the machinery behind stochastic optimization is briefly recalled, following \cite{Wechsung_2022, Wechsung_2}. The method involves generating multiple perturbed sample coil sets around the unperturbed sample, and then calculating the average field error for all samples. Averaging all the perturbed stellarators is supposed to smooth the descent path taken by the optimizer and result in wider local minima. The perturbation is a local and continuous displacement of the 3D filament of the coil, simulating real manufacturing deviations from the ideal model. In \texttt{SIMSOPT}, the perturbed path $\Tilde{\gamma_i}$ is produced by adding a Gaussian process (GP) \cite{Wechsung_2022} simulating a perturbation $\epsilon_i(l)$ to the unmodified coil path $\gamma_i(l)$, yielding a perturbed coil path $\Tilde{\gamma_i}(l)$, where $i$ refers to the coil number and $l$ to the arc-length index along the coil:

\begin{equation}
    \Tilde{\gamma_i}(l) = \gamma_i(l) + \epsilon_i(l) .
\end{equation}

A GP is a type of stochastic process and can be understood as a generalization of random samples with a multivariate normal distribution to functions. Here, the function is the perturbation and is indexed by the arc-length of the coil, and a discretization of the Gaussian process then follows at coil quadrature points. To produce the GP, a covariance matrix $\Sigma$ is computed, which sets the correlation between neighboring points of the perturbation. 

\begin{equation}
    \Sigma(l_i, l_j) = \tilde{k}(l_i, l_j) .
\end{equation}

In this work the calculation of the covariance matrix is performed via a well-behaved function of the radial basis function type:

\begin{equation}
    k(l_i, l_j) = \sigma^2 \exp{\left(\frac{-d^2}{2L_s^2}\right)} ,
\end{equation}

where $d=l_i-l_j$ is the distance between quadrature points, $\sigma>0$ is the amplitude of the perturbations in one dimension and $L_s>0$ determines the characteristic length of the oscillations. Moreover, a supplementary transformation from $k(d)$ to $\tilde{k}(d)$ enforces $2\pi$ periodicity \cite{Wechsung_2022}.

\begin{figure}[ht]
    \centering
    \includegraphics[scale = 0.12]{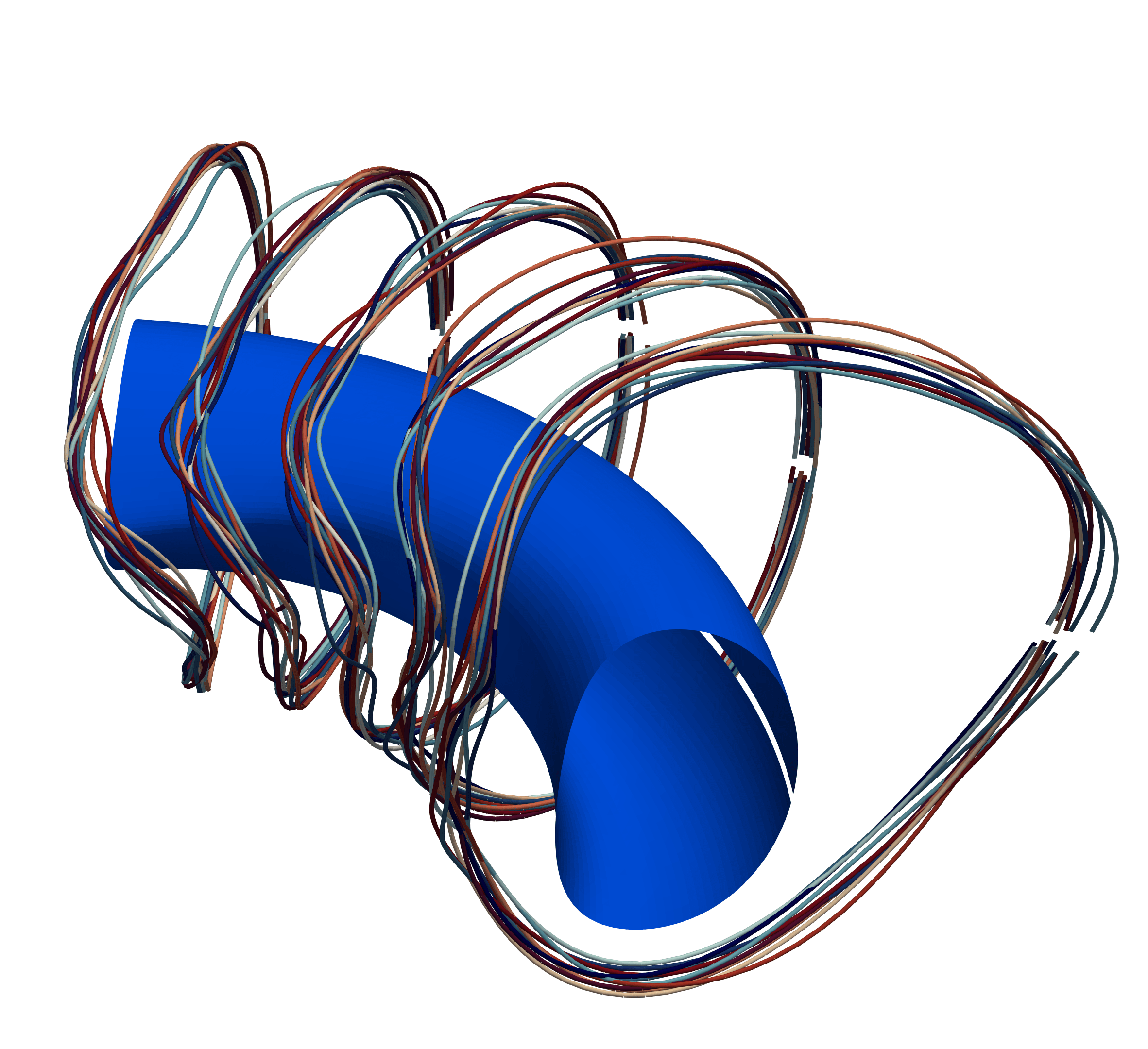}
    \caption{Illustration of 10 perturbed coils in a QA stellarator with 5-coils per half field period using the Gaussian Process approach. The blue surface is the last closed flux surface.}
    \label{fig:perturbed_samples}
\end{figure}

Finally, performing a Cholesky decomposition of the type: $\Sigma = DD^\top$, and then applying $D$ to a multivariate random vector $X$ with the same number of coefficients as quadrature points, yields the desired perturbed sample: $\epsilon_i = DX$. With multiple random vectors $X_i$, multiple samples are then drawn from a single coil set, where both the periodicity and amplitude of the perturbations can be tuned. In figure \ref{fig:perturbed_samples} an example of 10 samples for a QA stellarator show that the deviations appear continuous and smooth. In this example, the characteristic length $L_s$ was set to a fourth of the coil length ($\approx 0.3$m) and the amplitude $\sigma$ to 5 mm for visualization purposes.

Given a set of $N_{sample}$ samples of perturbed stellarators, the Biot-Savart law is applied $N_{sample}$ times, and the average field error is calculated:

\begin{equation}
    \langle f_{SF} \rangle = \frac{1}{N_{sample}} \sum_{i=1}^{N_{sample}} \frac{1}{2}\int_{S} \frac{\vert \Vec{B_i} \cdot \Vec{n}\vert^2}{\vert B_i \vert ^2}dS .
    \label{eq:stoch_squared_flux}
\end{equation}

This is a sample average approximation, as opposed to other risk-neutral or risk-averse stochastic optimization methods, for example. The expectation value in equation \ref{eq:stoch_squared_flux} replaces the deterministic squared flux in the optimization function. The minimizing algorithm is taken from scipy.minimize and employs the local optimizer BFGS~\cite{bfgs}. Note that the cloud of perturbed coils around the original unperturbed set is generated once in the initialization of the optimization routine, and is not recalculated at each optimization step; resampling is avoided as it appears to break down the BFGS approximation of the Hessian matrix. This can be explained by the fact that resampling modifies the objective function at every iteration, preventing a smooth evolution of the gradients.

\subsection{Optimization Workflow}
\label{sec:workflow_stoch}
To demonstrate the advantages of single-stage optimization, a method similar to that of Jorge et al. \cite{Jorge_2023} is implemented. The workflow is described below:

\begin{figure}[ht]
\centering
\includegraphics[width=0.6\linewidth]{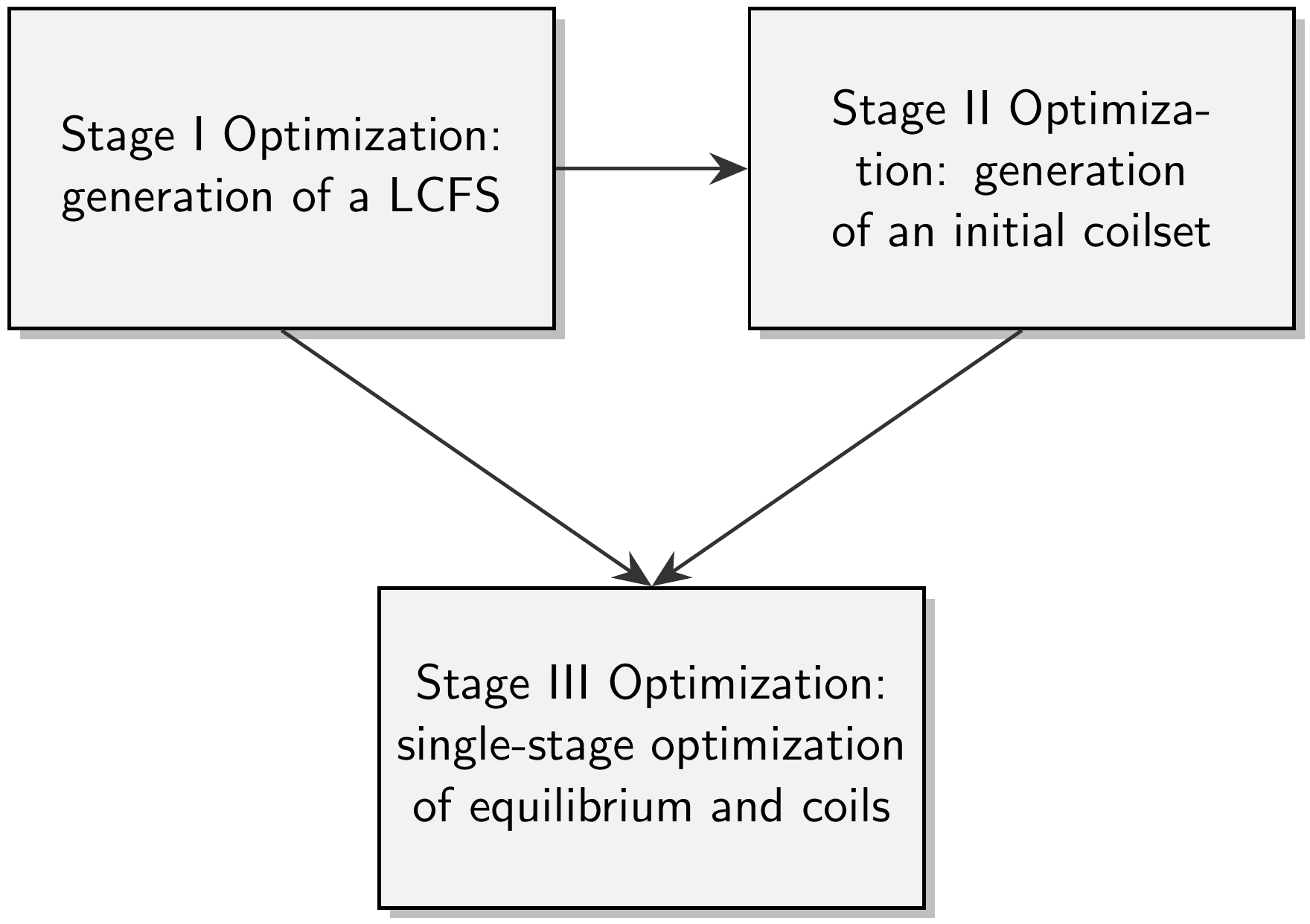}
\end{figure}

The single-stage process occurs after initial stage I and stage II optimizations are completed. In the first step, stage I optimization produces a partially optimized equilibrium for a "warm-start" for the single-stage optimization. Then that equilibrium is used as input for stage II optimization, where initial coils are found. Both the equilibrium from stage I and the coils from stage II are inputs to the single-stage optimization. 
The acceptability of a configuration and coil set is determined from the absolute value of the squared flux and the robustness of the given coil set. Greater robustness is achieved by avoiding sharp local minima, and instead finding flatter minima that allow for more deviation of the coils, as in stochastic optimization. 

To perform a posteriori assessments of the robustness of configurations, each coil set is perturbed $N_{sample}$ times. For each perturbed coil set, a squared flux minimizing (QFM) surface is found to describe the LCFS of the magnetic field generated by the coils. The QFM surface provides the boundary for fixed-boundary VMEC runs, which solve equilibria for each of the perturbed coil sets. All the robustness results presented are calculated from QFM surfaces, apart from when the squared flux metric is used.

Each optimization is performed on the VIPER cluster of MPCDF using 1280 samples, which corresponds to 5 nodes with 128 cores each, with a maximum required RAM usage of 200 GB. One run lasts on average $\simeq 1$h, depending on the complexity of the configuration. The hardware consisted of AMD EPYC Genoa 9554 CPUs with at least 512 GB RAM per node.

\section{Results}
\subsection{Parameter Tuning and Parallelization}

\begin{figure}[ht]
    \centering
    \includegraphics[width=0.45\linewidth]{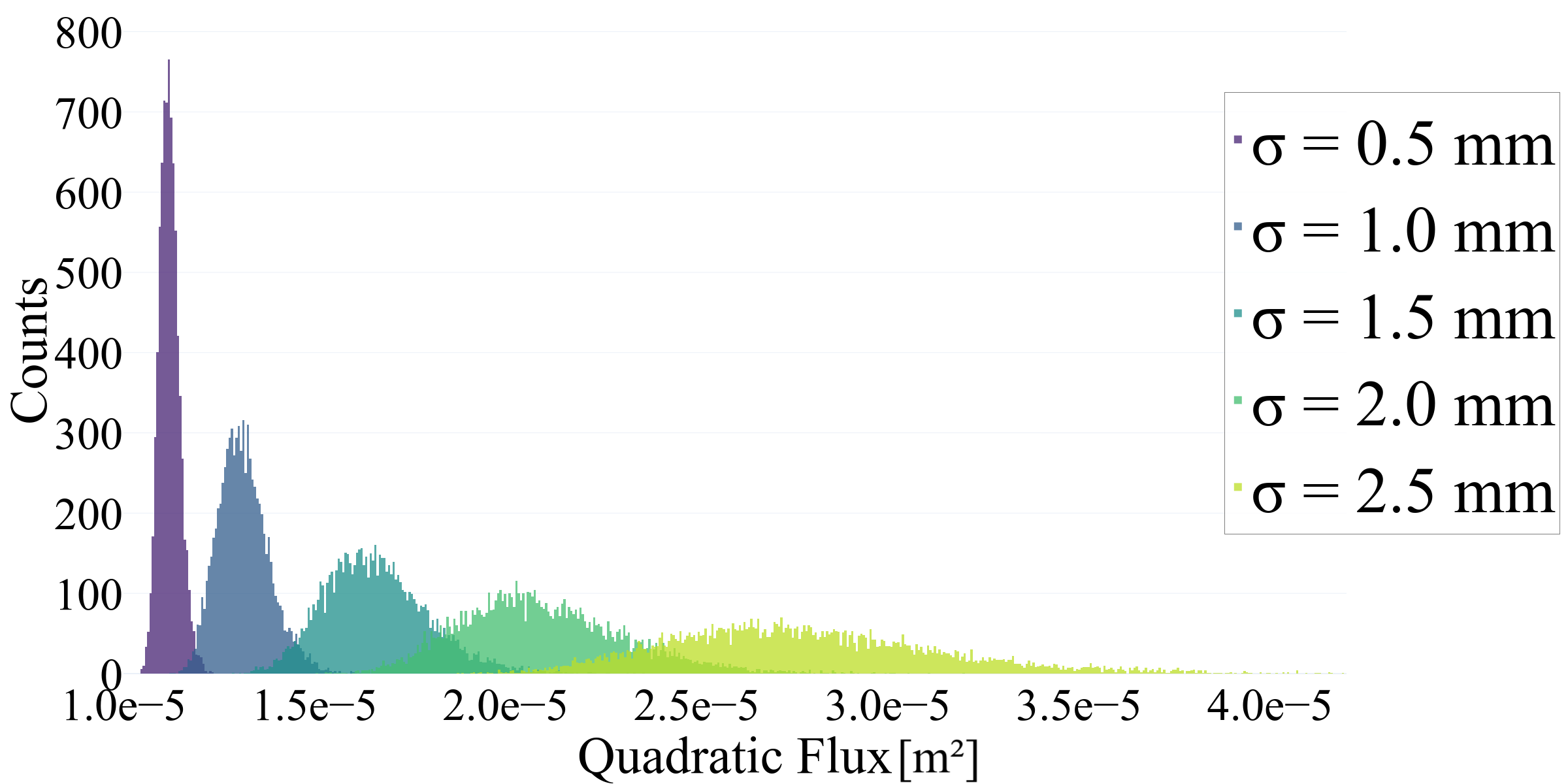}
    \includegraphics[width=0.45\linewidth]{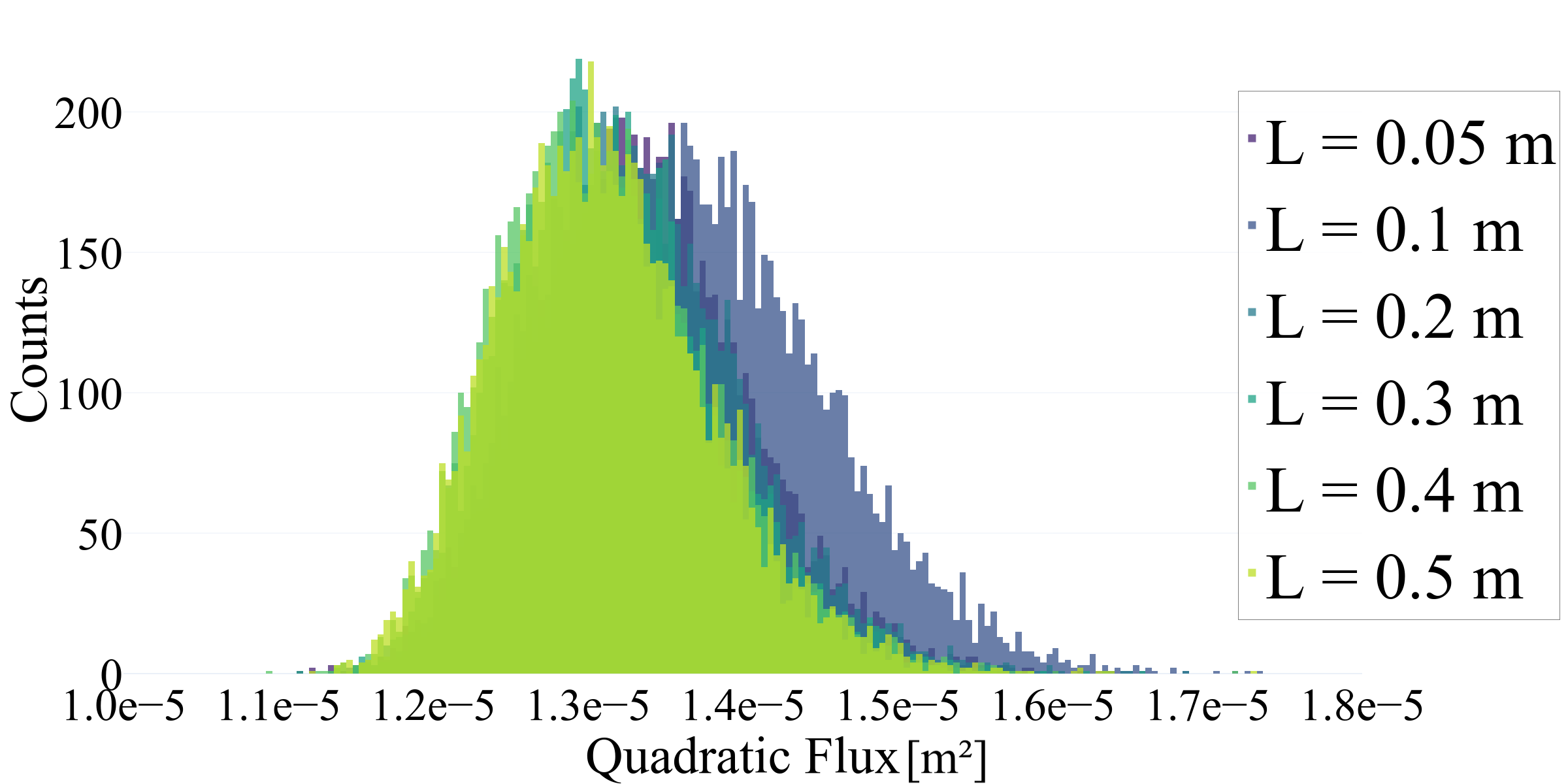}
    \caption{\textbf{Left}: Distributions of the squared flux for different coil perturbation amplitudes $\sigma$ for a HSX-like 4 field period QH stellarator with 1 m major radius. \textbf{Right}: Distributions of the squared flux for different coil perturbation wavelengths $L$.}
    \label{fig:qa_distributions}
\end{figure}

The inherent geometry of the coils can contribute to a higher robustness of the equilibrium, i.e certain geometries can generate magnetic fields that are less sensitive to perturbations. (\cite{Landreman_2018_shape_gradient, Paul_Antonsen_Landreman_Cooper_2020_shape_gradient, Antonsen_Paul_Landreman_2019_shape_gradient}). Selecting the correct perturbation amplitude and wavelength, $\sigma$ and $L$ respectively, is important to achieving both a robust and precise configuration. Shown on the left of Figure \ref{fig:qa_distributions} are distributions of the squared flux for 8,000 perturbations of an HSX-like 4 field period QH stellarator at values of $\sigma$ between 0.5 mm and 2.5 mm. The distribution is right-skewed, which indicates that random perturbations tend to deteriorate the original stellarator accuracy more than improve it at fixed $\sigma$. This distribution will increase the average squared flux. For a 1 m major radius device, deviations of 2.5 mm degrade the magnetic field accuracy by about a factor of 4. which means that, for a given robustness and manufacturing accuracy, any extra refinement of the plasma configuration coming from optimization is lost to perturbations. Over-optimizing both quasisymmetry (e.g. $f_{QS}<10^{-3}$) requiring field accuracies to magnitudes of around $\langle B\cdot n\rangle / \langle B \rangle< 10^{-3}$ does not benefit a device if built accuracies are 2 mm (these can deteriorate approximately the field accuracy by $10^{-2}$ \cite{andreeva_2004, Andreeva_2015}). Distributions of the squared flux for values of the characteristic length L between [0.2 m - 0.7 m] are shown on the right of Figure \ref{fig:qa_distributions}. With a major radius of 1 m, it can be concluded that the choice of L does not impact the field quality significantly compared to increasing the amplitude. Determining the parameters $\sigma$ and $L$ to simulate real-world coil deformations is largely dependent on the manufacturing process \cite{gil2025manufacturingtolerancesnonplanarcoils}, which may not be the same as the optimal parameters for efficient single-stage convergence.

\begin{figure}[ht]
    \centering
    \includegraphics[width=0.47\linewidth]{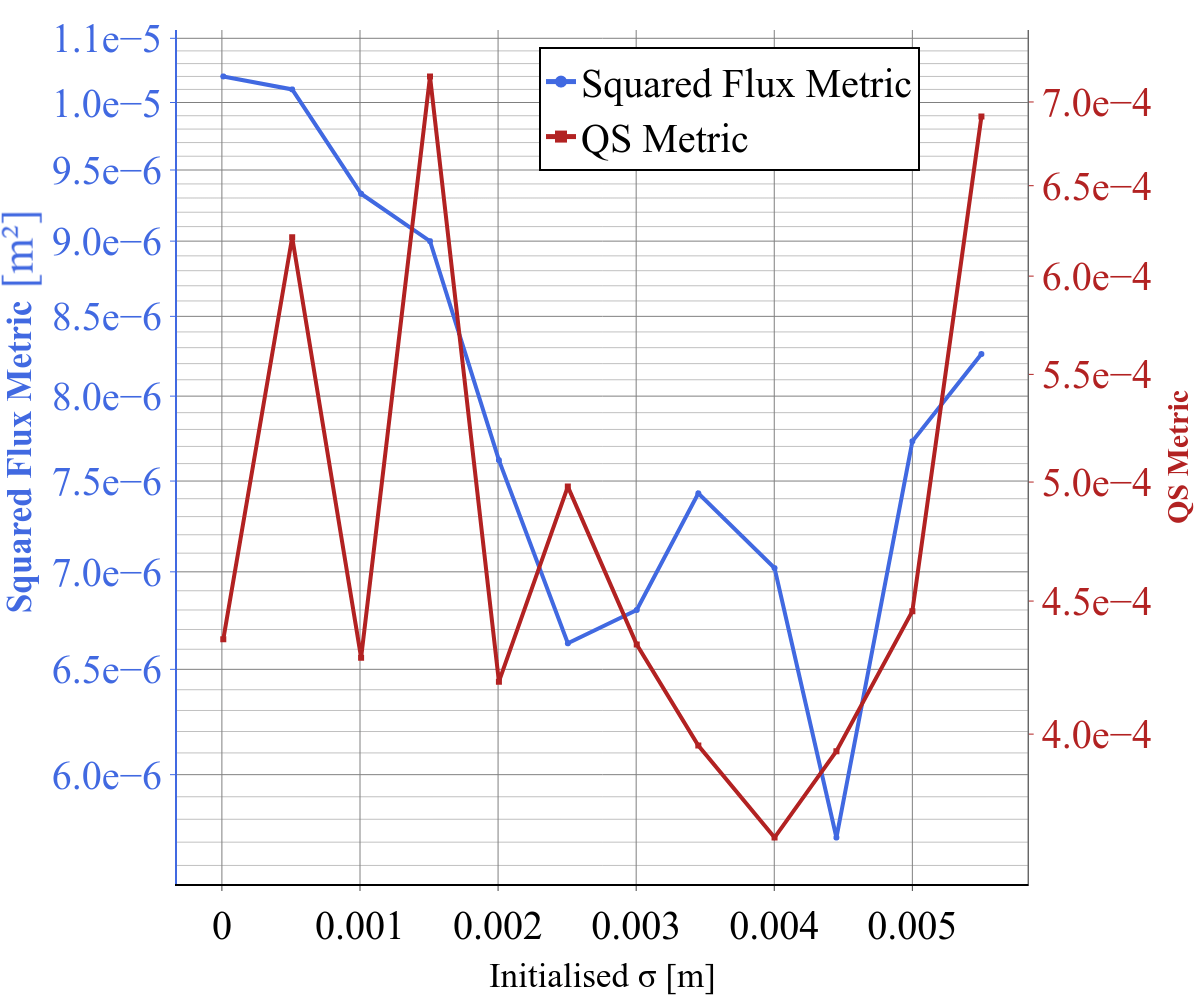}
    \includegraphics[width=0.47\linewidth]{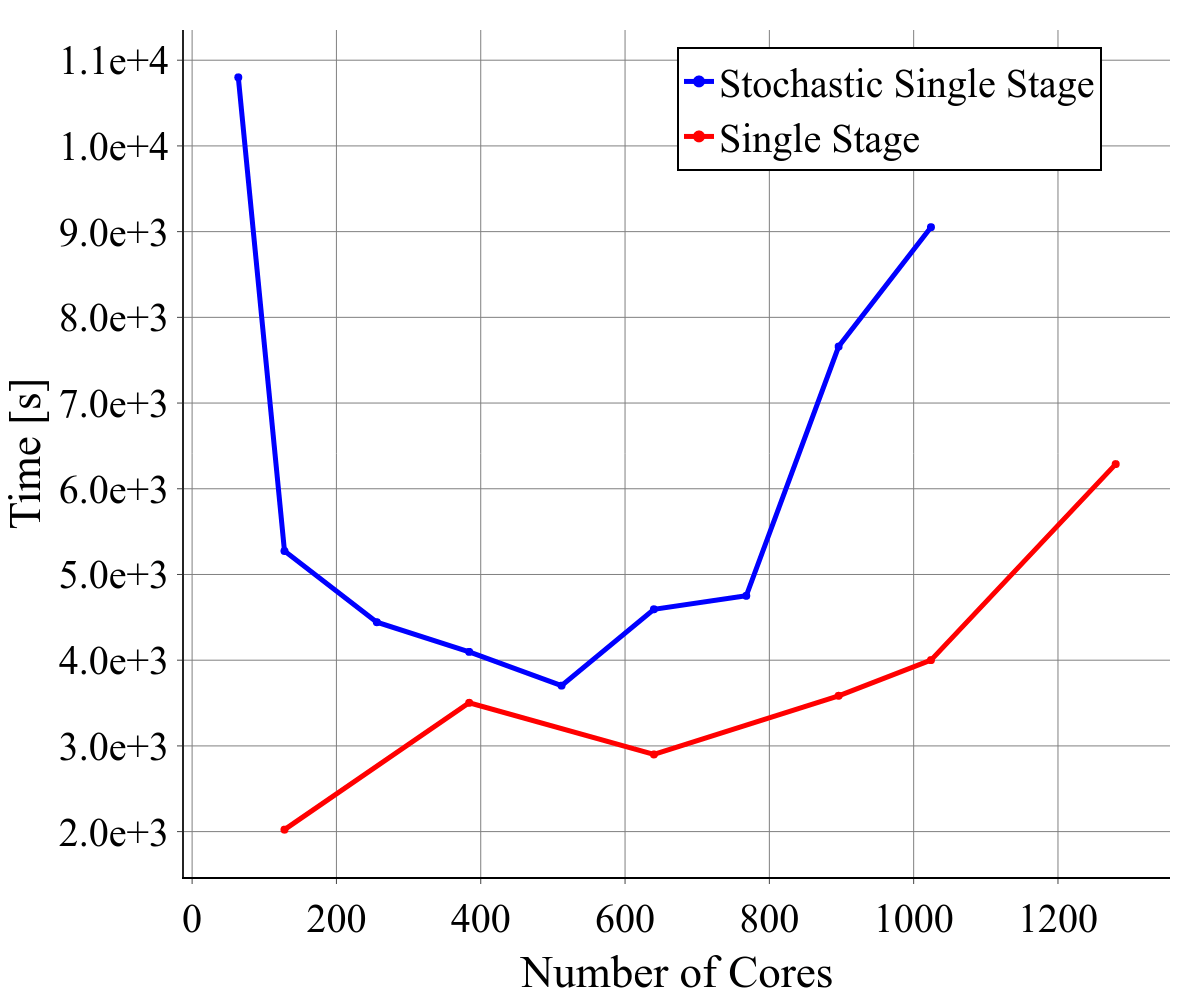}
    \caption{\textbf{Left}: Quasisymmetry measured from the final optimized fixed-boundary equilibrium object as a function of the initialized perturbation amplitude $\sigma$. \textbf{Right}: Run time versus number of cores used. Between 64 and 512 cores, the runtime decreases as expected due to parallelization for the stochastic single-stage (blue). However, after this point, concurrency issues, most likely arising from the standard single-stage run (red), appear and delay the optimization.}
    \label{fig:sigma_opt_impact}
\end{figure}

 Furthermore, in Figure \ref{fig:sigma_opt_impact}, two optimization metrics are plotted: the quasisymmetric error and the squared flux after stochastic single-stage optimization for different initial $\sigma$ values. Both metrics appear to progressively reach lower minima as the perturbation amplitude of the coils increases. Intuitively, the optimized squared flux improves because increasing $\sigma$ results in higher average values for the objective function, as is shown in Figure \ref{fig:qa_distributions}, and the optimizer interprets it as an increase in the hyperparameter governing the squared flux. This, however, should compete with the QS error and prevent its efficient minimization. One hypothesis for the decrease in the quasisymmetric error is that the optimizer becomes less stuck in minima for the coils, which indirectly impacts the equilibrium optimization. 

When executing Monte-Carlo simulations or any work that requires a considerable number of samples, it is usually beneficial to parallelize as much as possible for increased numerical efficiency. In our case, one of the main hurdles preventing fast optimization is the estimation of the average Jacobian from N samples while making it compatible with the estimation of the surface component $\partial F_S/\partial \textbf{x}_S$  (itself already parallelized). The standard single-stage optimization process is as follows: processor 0 (main) performs the optimization. At the same time, the calculation of each component of the Jacobian is delegated to each of the other processors. For the standard stochastic stage II, optimization is performed by every processor, but each processor evaluates only one squared flux sample. This is then propagated to all the other processors, which in turn estimate the average. The incompatibility arises from the fact that the standard single-stage only has one processor performing the optimization, while stochastic optimization requires all of them to perform the optimization. In this work, we corrected this incompatibility by imposing that every processor must perform the same optimization and propagate both the squared flux value of its own sample and its Jacobian component of $\partial F_S/\partial \textbf{x}_S$ to all the other processors. This results in every unit having the surface Jacobian and average squared flux available to proceed with the Hessian estimation. To verify that parallelization works, we progressively increase the number of cores and measure the runtime. As shown on the right graph from Figure \ref{fig:sigma_opt_impact}, there
is an initial regime where run time decreases as expected; however, the trend reverses at around 512 cores. This is most likely due to concurrency issues originating from the standard single-stage code, where increasing the core number prevents efficient decomposition of the Jacobian. This can come into conflict with the squared flux sampling parallelization. However, there is a region between 128 and 512 cores, where the run time appears optimal. These numbers are also dependent on the number of cores per node. These runs are performed on the MPCDF Viper Cluster, which features 128 cores per node.
 
\subsection{Optimization Path}

\begin{figure}[ht]
    \centering
    \includegraphics[width=0.8\linewidth]{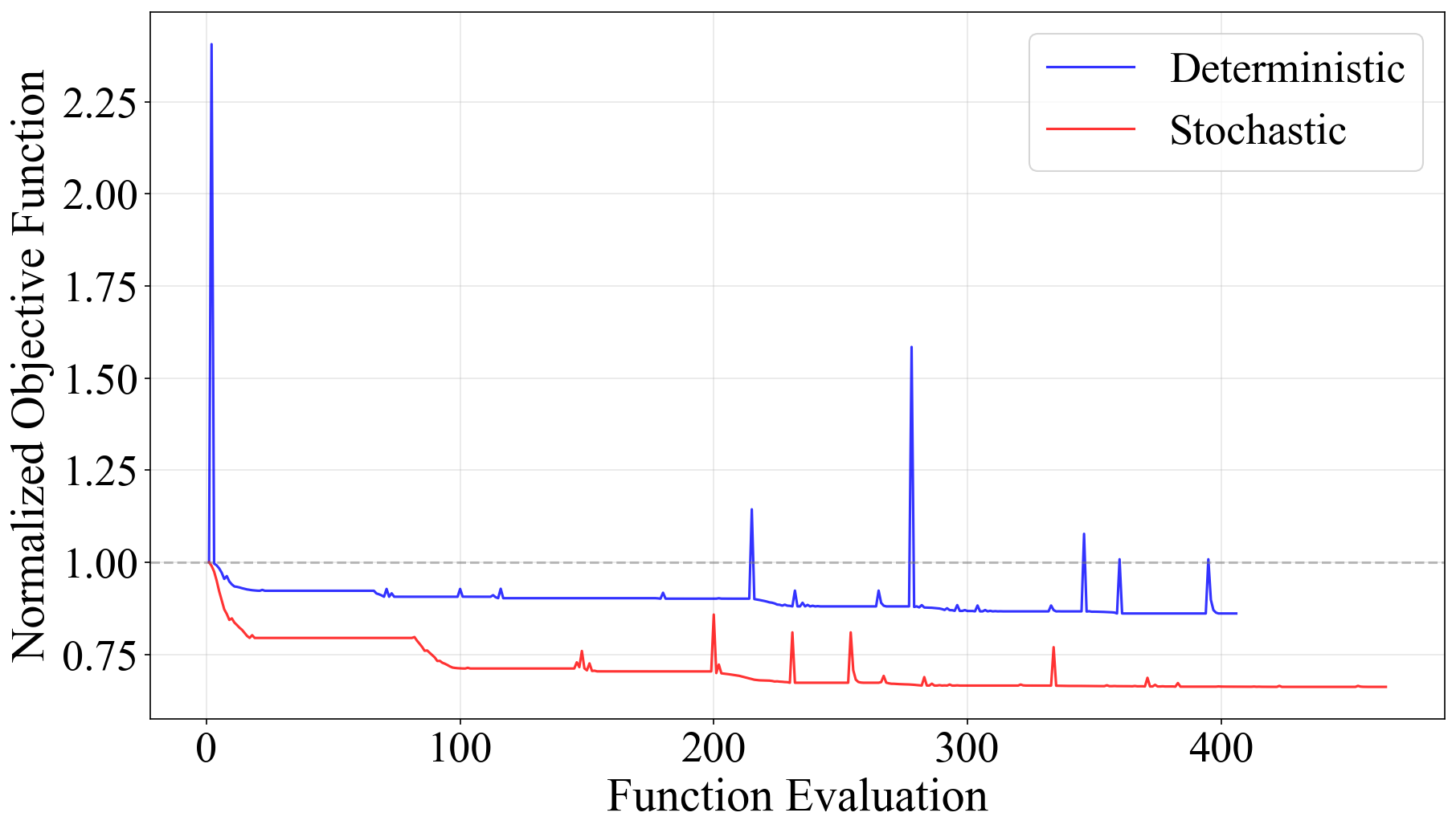}
    \caption{Total minimizing function during the optimization for both standard single-stage (blue) and stochastic single-stage (red). Results are normalized with regards to the respective initial value for comparison. The stochastic single-stage approach appears to not get stuck earlier during the optimization allowing the minimizer to reach deeper minima.}
    \label{fig:convergence_plot}
\end{figure}

To characterize the performance of the proposed methodology, Figure \ref{fig:convergence_plot} shows the evolution of the normalized objective function as a function of the number of iterations for both the deterministic and stochastic single-stage optimization frameworks. The initial phase of the optimization for both methods shows a rapid descent, indicating an efficient navigation of the high-dimensional parameter space. While the deterministic optimization (blue curve) successfully reduces the objective function below its initial value, the convergence eventually plateaus at a normalized value of approximately 0.85, suggesting it has reached a local minimum sensitive to small variations in the configuration. In contrast, the stochastic optimization (red curve) demonstrates a more stable and deeper convergence profile. By targeting a "cloud" of perturbed coil configurations rather than a single discrete geometry, the stochastic framework effectively smoothens the objective landscape, prioritizing broader minima. This is visible in the lower asymptotic value achieved by the stochastic method, which settles near 0.65. The reduction in both the frequency and magnitude of the spikes compared to the deterministic case suggests that the stochastic objective function provides a more well-behaved landscape for the optimizer. Ultimately, the stochastic single-stage approach not only yields a configuration with a lower objective value but should also ensures that the resulting stellarator design remains high-performing even in the presence of the geometric uncertainties discussed in Section \ref{sec:optimization_methods}.

\label{sec:numerical_results}
\subsection{Quasi-axisymmetric stellarator}

As a first application of this method, a 1 m major radius, three-field-period NCSX-like QA stellarator is chosen. The optimization is performed following the workflow described in Section \ref{sec:workflow_stoch}. The properties of the initial condition that is fed into the stochastic single-stage loop are given in Table \ref{tab:warm_start_properties}. The coils are optimized to a relatively high precision in the initial stage II, and similarly, the equilibrium yields low quasi-symmetric errors ($< 1\times 10^{-3}$). This very warm-start is intentionally made, as the goal of a stage III optimization is to modify both surfaces and coils slightly to find a more compatible/robust configuration. 

\begin{table}[ht]
\centering
\begin{tabular}{|c|c|}
\hline
Coil and Eq. Properties                     & Warm-start QA \\ \hline
$\langle B\cdot n\rangle / \langle B \rangle$ &   1.39e-03                                \\ \hline
QS error (Fixed-Boundary) [a.u]                             & 4.75e-4                              \\ \hline
Total Coil Length [m]                               & 13.21                                    \\ \hline
Mean $\iota$                        &  0.4                                            \\ \hline
Aspect Ratio                        &  6                                            \\ \hline
\end{tabular}
\caption{Coil and equilibrium parameters of the warm-start configuration used for the QA stellarator. Here, the total coil length designates the total coil length per half-field period of the stellarator.}
\label{tab:warm_start_properties}
\end{table}

The stochastic single-stage optimization performed for this configuration serves as a first demonstration; therefore, coil cost functions are limited to the coil length and stochastic squared flux. The coils are optimized to have at most 0 length, i.e. coil length is minimized. Regarding the equilibrium constraints, precise quasisymmetry across all surfaces, as well as an iota of 0.4 and an aspect ratio of 6, are targeted. The values for the rotational transform and aspect ratio are identical to those of the stage I optimization. This way, they prevent the optimizer from findng unwanted minima by generating equilibria with a higher aspect ratio and a lower iota that possess very low levels of quasisymmetric errors. To benchmark the performance of the stochastic single-stage optimization, three other optimizations are performed in parallel. Two of them are performed with the standard form of single-stage as presented in \cite{Jorge_2023}: QA Std. L-OFF is a single-stage optimization without the length constraint and QA Std. L-ON is done with the length constraint. The relaxation of the length constraint is done in order to restrain the optimizer as little as possible. The third optimization is chosen to be a standard stage I / stochastic stage II optimization, here labeled as QA Stoch SII. All stochastic optimizations are executed with $\sigma = 2.5$ mm and $L=0.2$ m initial coil perturbation parameters. The dynamic resolution of the surface modes is performed from mode one to mode 12, as it corresponds to three times the number of coils per half-field period, allowing for the resolution of magnetic field ripples from the discrete number of coils.

The reasoning behind the benchmark is as follows: the standard optimization versions serve as checks on both the performance of the stochastic minimization version of the code and on the effectiveness of stochastic optimization in increasing stellarator robustness. Note that for all the single-stage optimization, the same hyperparameters are used everywhere. The Stoch SII optimization serves as a check to verify if similar levels of robustness are achieved as those of an already verified method. These hyperparameters were determined to yield similar field accuracy, allowing for a fair robustness comparison. 

The robustness checks are performed a posteriori to the optimization, and they consist of generating $N_s$ QFM samples, from $N_s$ different perturbed stellarators for increasing $\sigma$. This corresponds to an additional step compared to \cite{Wechsung_2}, as it is an equivalent free-boundary robustness check for the equilibria and does not rely solely on the squared flux. For the QA stellarator, the sampling is done with N = 128, and $\sigma$ is chosen between 0 and 3 mm, and $L = 1$ m. In order to be able to resolve coil ripple, each QFM surface is optimized with M = 18 and N = 30 (see Equation \ref{eq:surface}). 

\begin{table}[ht]
\centering
\begin{tabular}{|c|c|c|c|c|}
\hline
    & Std. L-OFF & Std. L-ON  & Stoch SII & Stoch L-ON \\ \hline
Max $(B\cdot n) /  B $ &                  5.43e-3                            &                           5.96e-3               &         5.62e-3           &      \textbf{4.74e-3}      \\ \hline
$\langle B\cdot n\rangle / \langle B \rangle$ &           1.17e-3                                  &                    1.37e-3                       &           1.01e-3         &      \textbf{9.89e-4}      \\ \hline
Total Coil Length [m]                    &    \textbf{13.44}     &      13.99          &       14.0          &    13.99       \\ \hline
CS distance [m]                       &   0.178      &   0.19     &    \textbf{0.2}       &     0.19                 \\ \hline
CC distance [m]                       &    \textbf{0.08}     &    0.038         &      0.02         &  0.038            \\ \hline
Squared Flux metric [m$^{2}$]                   &    7.69e-6       &   8.06e-6       &   6.37e-6      &      \textbf{6.35e-06}          \\ \hline
QS error (Fixed-Boundary) [a.u]                   &    4.15e-4       &     4.22e-4      &   4.75e-4*      &      \textbf{3.94e-4}          \\ \hline
QS error (QFM) [a.u]                   &    \textbf{1.2e-3}       &     1.3e-3      &    1.3e-3      &      1.3e-3          \\ \hline
QS at $\sigma =$ 3 mm   [a.u.]           &   1.6e-2      &     1.6e-2        &      \textbf{1.2e-2}            &   1.3e-2          \\ \hline
Average iota & 0.40 & 0.40 & 0.40* & 0.40 \\ \hline 
Aspect Ratio & 6.04 & 6.03 & 6.00* & 6.04\\ \hline 
\end{tabular}
\caption{Properties of both QA coils and the equilibria that are given to the optimizer for the NCSX-like stellarator in the four different optimizations. Here we compare the stochastic single-stage (Stoch L-ON) to both traditional stage I and stochastic stage II optimization (Stoch SII) and standard single-stage (Std. L-OFF \& Std. L-ON) according to \cite{Jorge_2023}. QS at $\sigma=3$ mm refers to an average quasisymmetric error measured for samples perturbed with a coil deviation amplitude of about 3 mm. The Squared Flux metric and QS error in fixed-boundary correspond to the values given by the optimizer at the end of the optimizations. *These values correspond to the values of the warm-start stage I equilibrium shown in Table \ref{tab:warm_start_properties}.}
 \label{tab:qa_opts}
\end{table}

\begin{figure}[ht]
	\centering
	\includegraphics[width=0.47\linewidth]{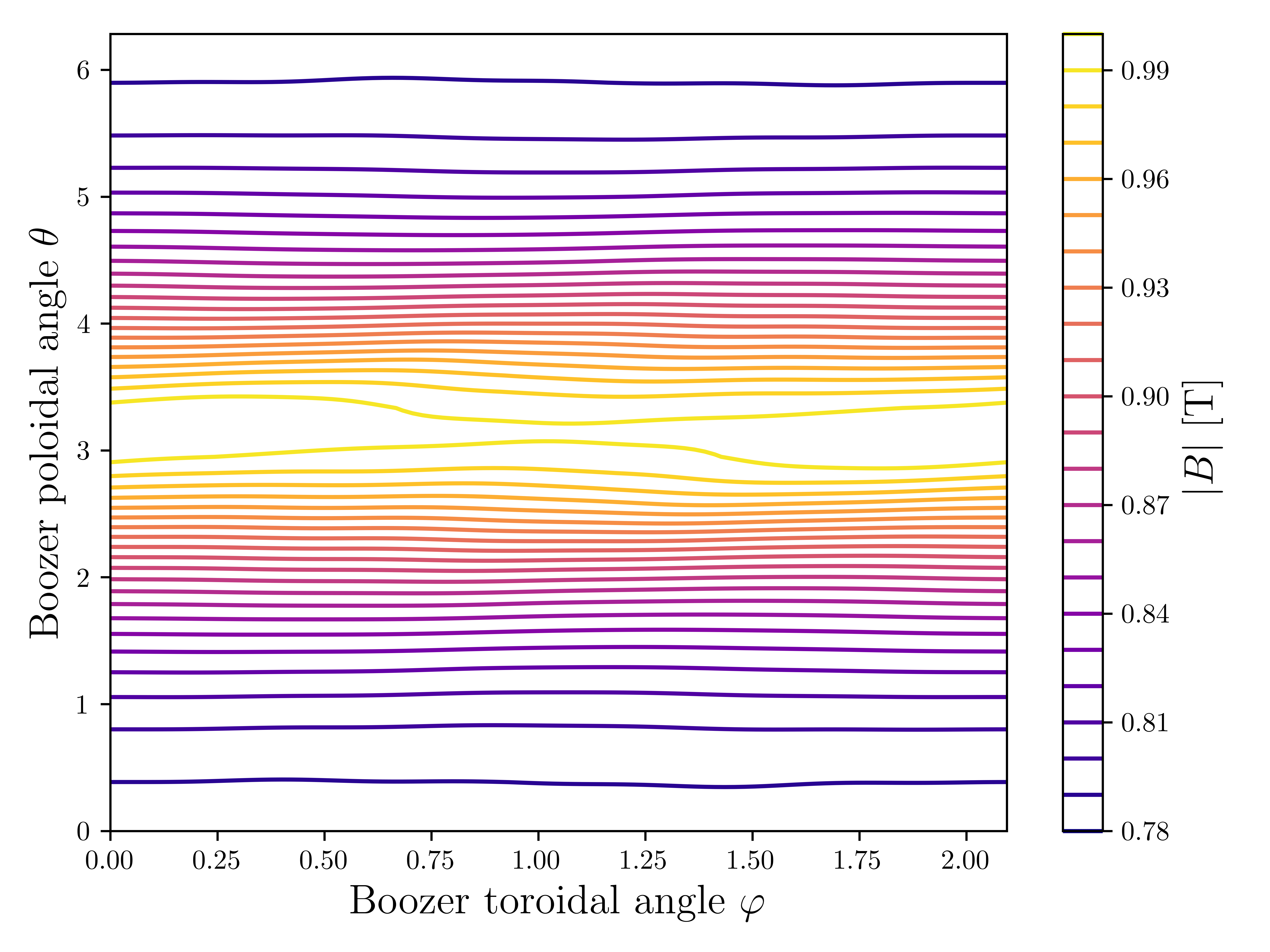}
	\includegraphics[width=0.47\linewidth]{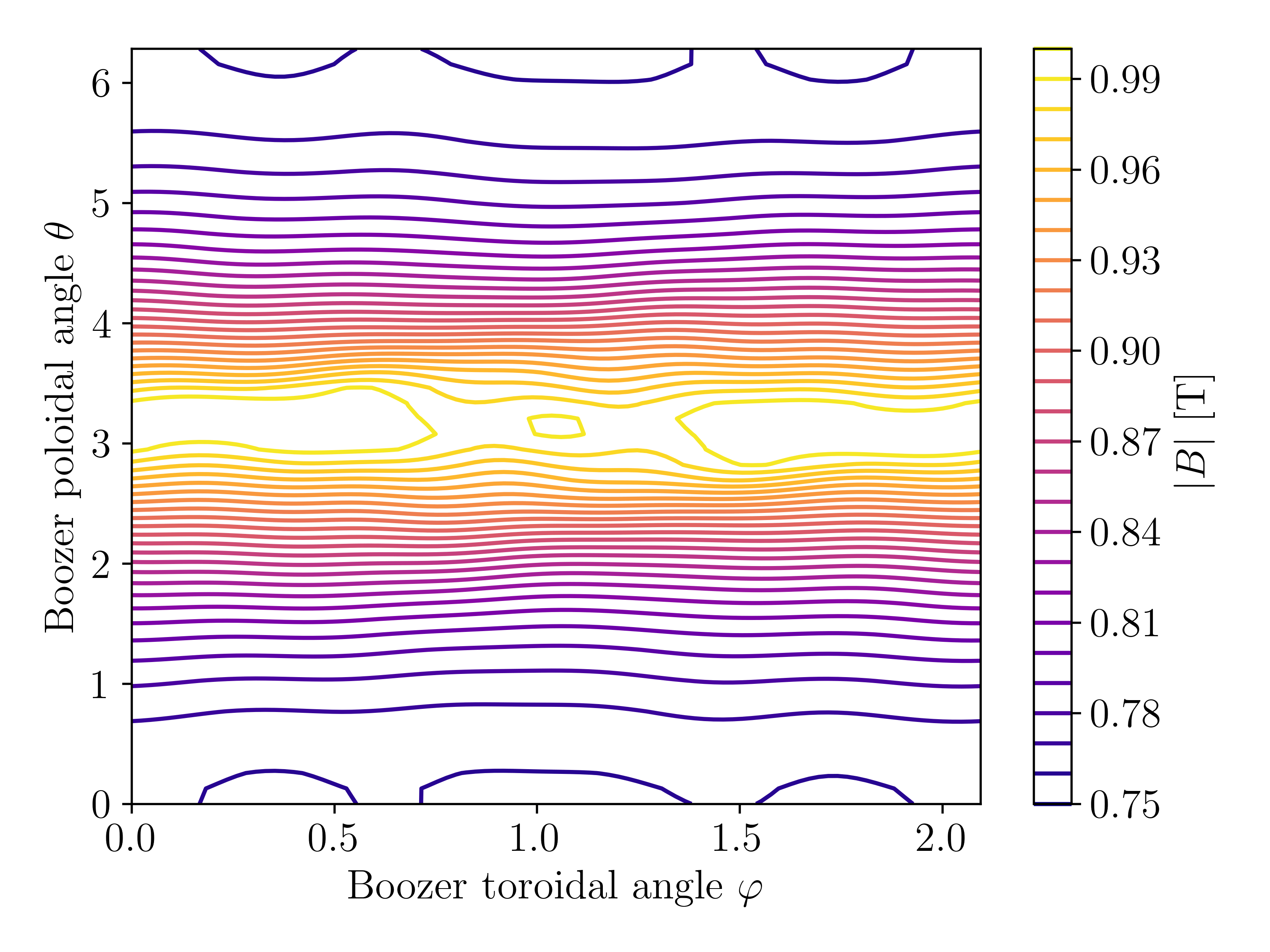}
	\caption{\textbf{Left:} Boozer plot at s=0.9 displaying qualitatively the departure from quasisymmetry from the stochastic single stage QA equilibrium calculated using QFM surfaces. No coil ripple is visible, and mostly straight contours are visible, showcasing an example of good quasisymmetry. \textbf{Right:} Boozer plot at s=0.9 of the same coil set but perturbed with $\sigma = 3$ mm. Despite visible perturbations, quasisymmetry is retained.}
	\label{fig:qa_boozer}
\end{figure}

The results of the four different optimizations are displayed in Table \ref{tab:qa_opts}. The stochastic single-stage optimization appears to outperform the other comparisons in terms of average and maximum field accuracy, as well as quasisymmetry achieved in the fixed-boundary optimization. In Figure \ref{fig:qa_boozer}, the unperturbed Boozer plot and one example of perturbation of the field at $\sigma=3$ mm are shown, displaying the degree of quasisymmetry achieved to a relatively high degree in both cases. 

The improvements from the equivalent stage III optimizations presented here are relatively minor compared to the initial conditions. Note that the Std. L-ON run made barely any improvement on the squared flux and quasisymmetry. This is in part due to the very warm-start, but also partially due to the difficulty for the optimizer to leave already good local minima. The results presented in the work of Jorge et al. \cite{Jorge_2023} for the four-coil QA stellarator case appear more optimistic since the initial conditions were relatively worse, with the initial stage II yielding coils with a squared flux metric of 6.3e-4 $\text{m}^{2}$. Here, a warmer start is also chosen to investigate the potential of stochastic methods to depart from local minima. While the improvements appear marginal in terms of performance compared to the original method, differences become apparent when considering robustness. The stochastic approach seems to meet similar performance as a stochastic stage II concerning the robustness of the stellarator as it achieves a reduction of about 19 \% in quasisymmetric error as defined in Equation \ref{eq:quasisymmetry}.

\begin{figure}[ht]
	\centering
	
	
	\begin{subfigure}[t]{0.35\textwidth}
		\includegraphics[width=\textwidth]{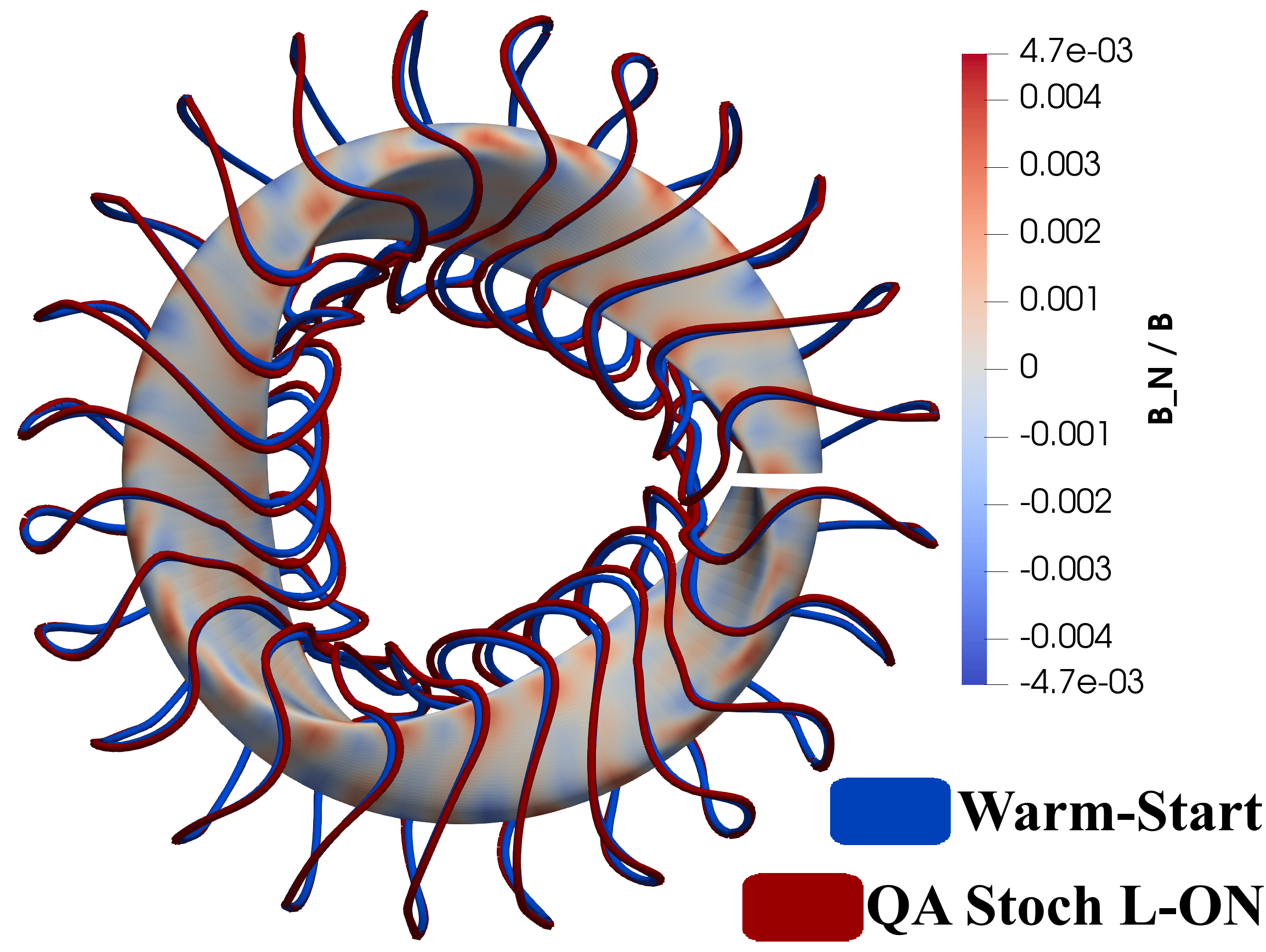}
		\caption{}
		\label{fig:qa_coils_warm_start_stoch}
	\end{subfigure}
	\hspace{1cm}
	\begin{subfigure}[t]{0.35\textwidth}
		\includegraphics[width=\textwidth]{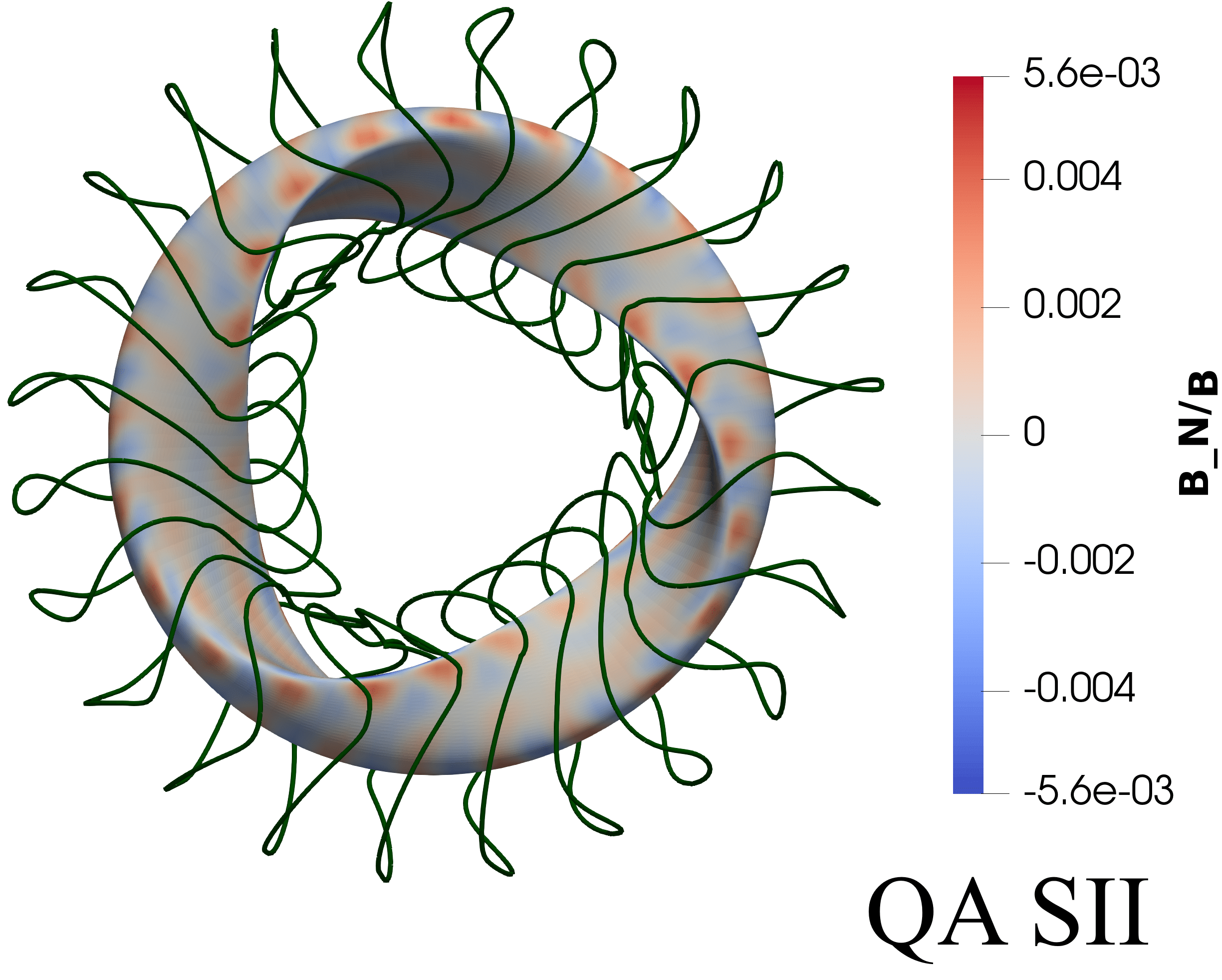}
		\caption{}
		\label{fig:coils_qa_SII}
	\end{subfigure}
	
	\par\bigskip
	
	\begin{subfigure}[t]{0.35\textwidth}
		\includegraphics[width=\textwidth]{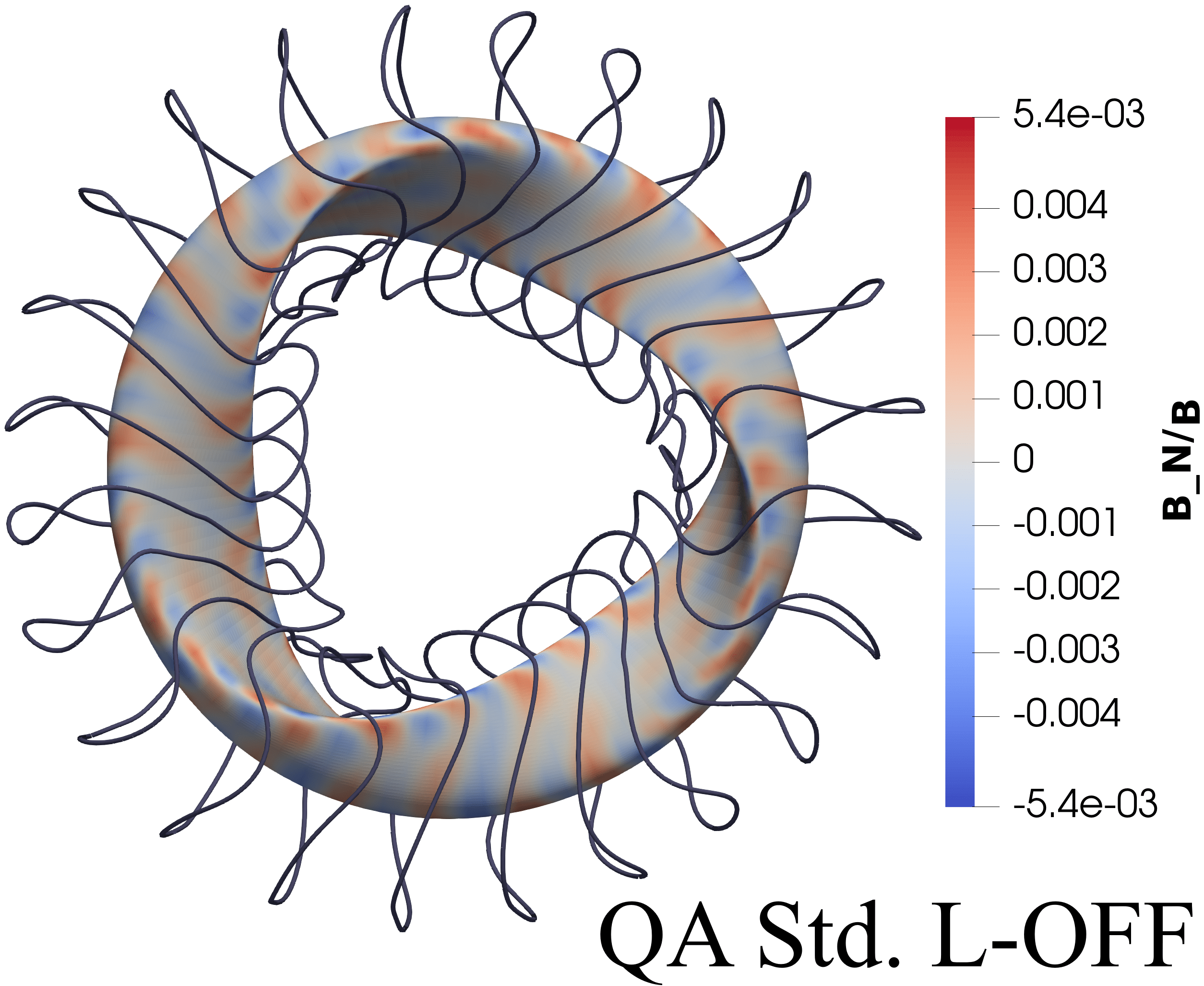}
		\caption{}
		\label{fig:coils_qa_single_loff}
	\end{subfigure}
	\hspace{1cm}
	\begin{subfigure}[t]{0.35\textwidth}
		\includegraphics[width=\textwidth]{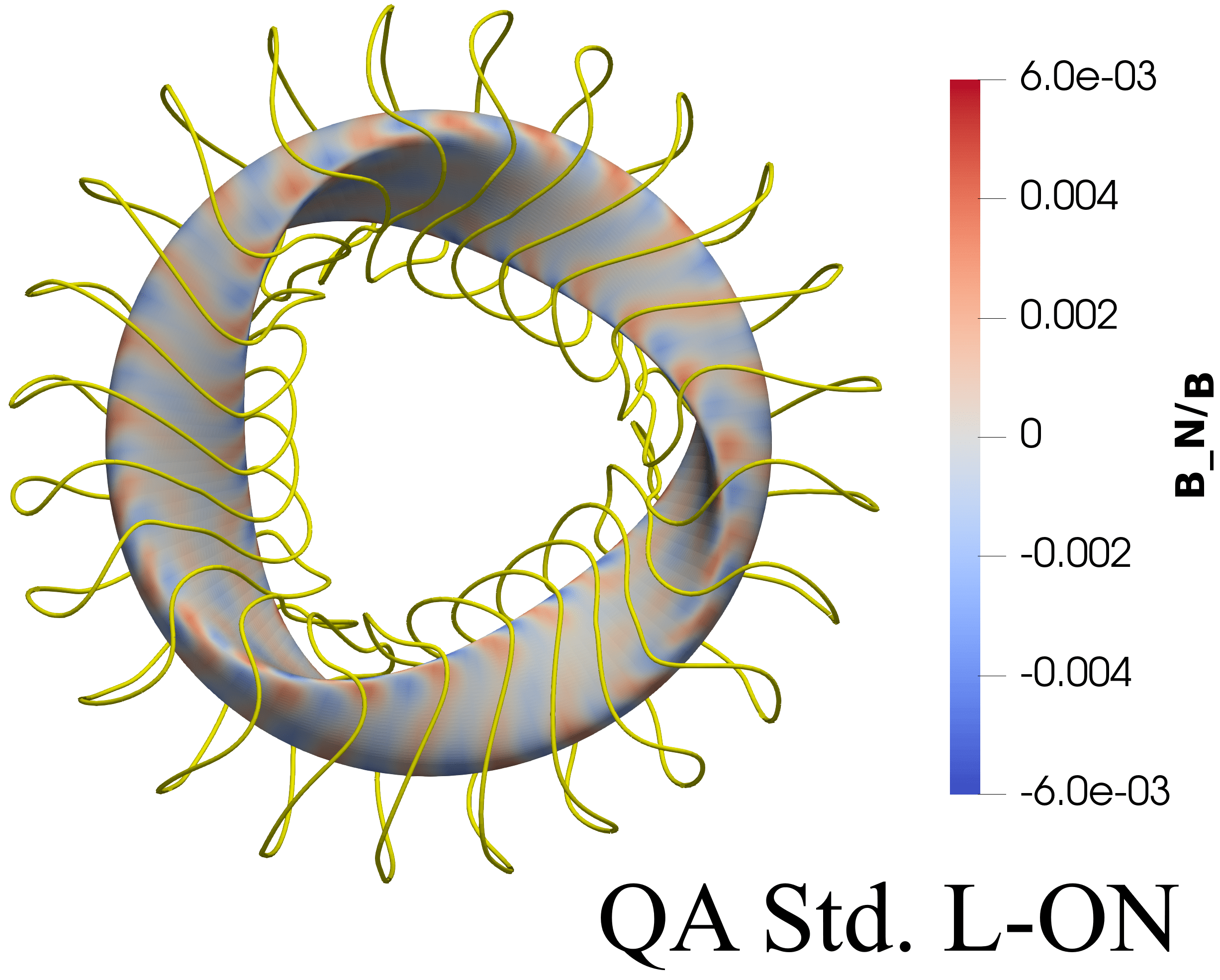}
		\caption{}
		\label{fig:coils_qa_single_lon}
	\end{subfigure}
	
	\caption{ Results of the stochastic single stage method applied to the quasi-axisymmetric case. Coils obtained from a stage II warm-start and stochastic single stage (a), coils obtained from stochastic stage II (b),  coils obtained from standard single stage without coil length penalty (c), coils obtained from standard single stage with coil length penalty (d).}
	\label{fig:qa_coils_all}
\end{figure}

\begin{figure}[ht]
	\centering
	
	
	\begin{subfigure}[t]{0.4\textwidth}
		\includegraphics[width=\textwidth]{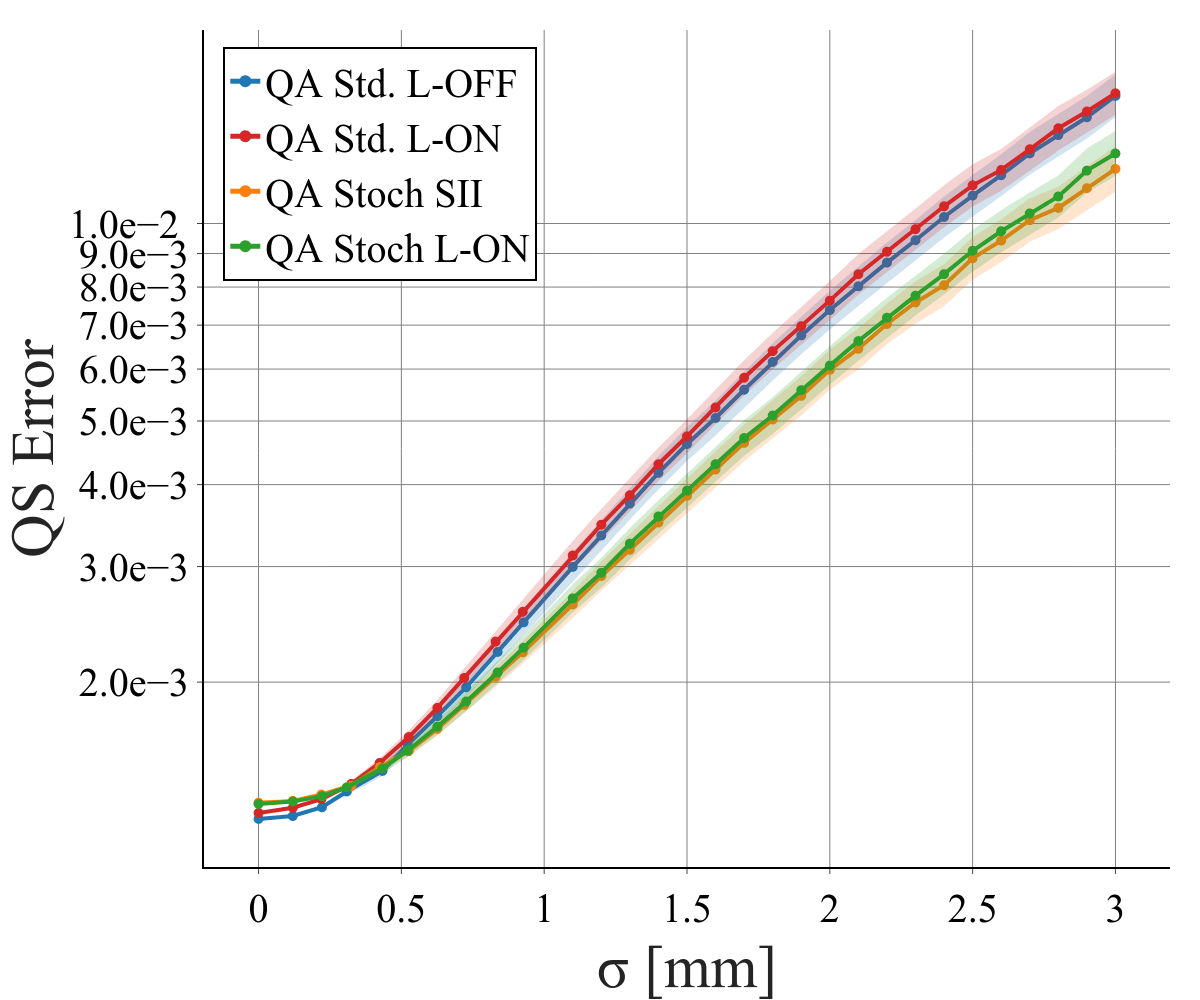}
		\caption{}
		\label{fig:a}
	\end{subfigure}
	\begin{subfigure}[t]{0.4\textwidth}
		\includegraphics[width=\textwidth]{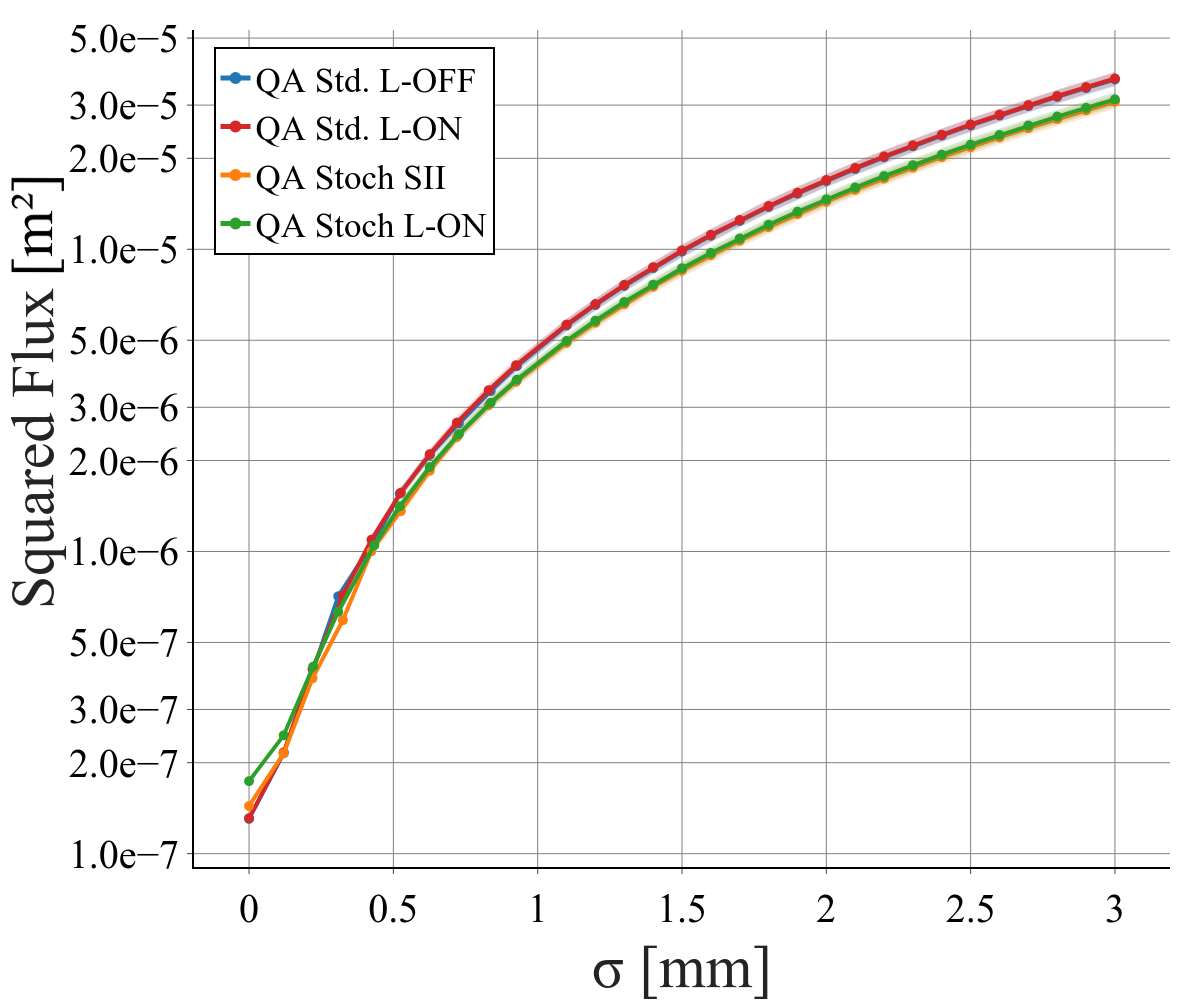}
		\caption{}
		\label{fig:b}
	\end{subfigure}
	
	\par\bigskip
	
	\begin{subfigure}[t]{0.4\textwidth}
		\includegraphics[width=\textwidth]{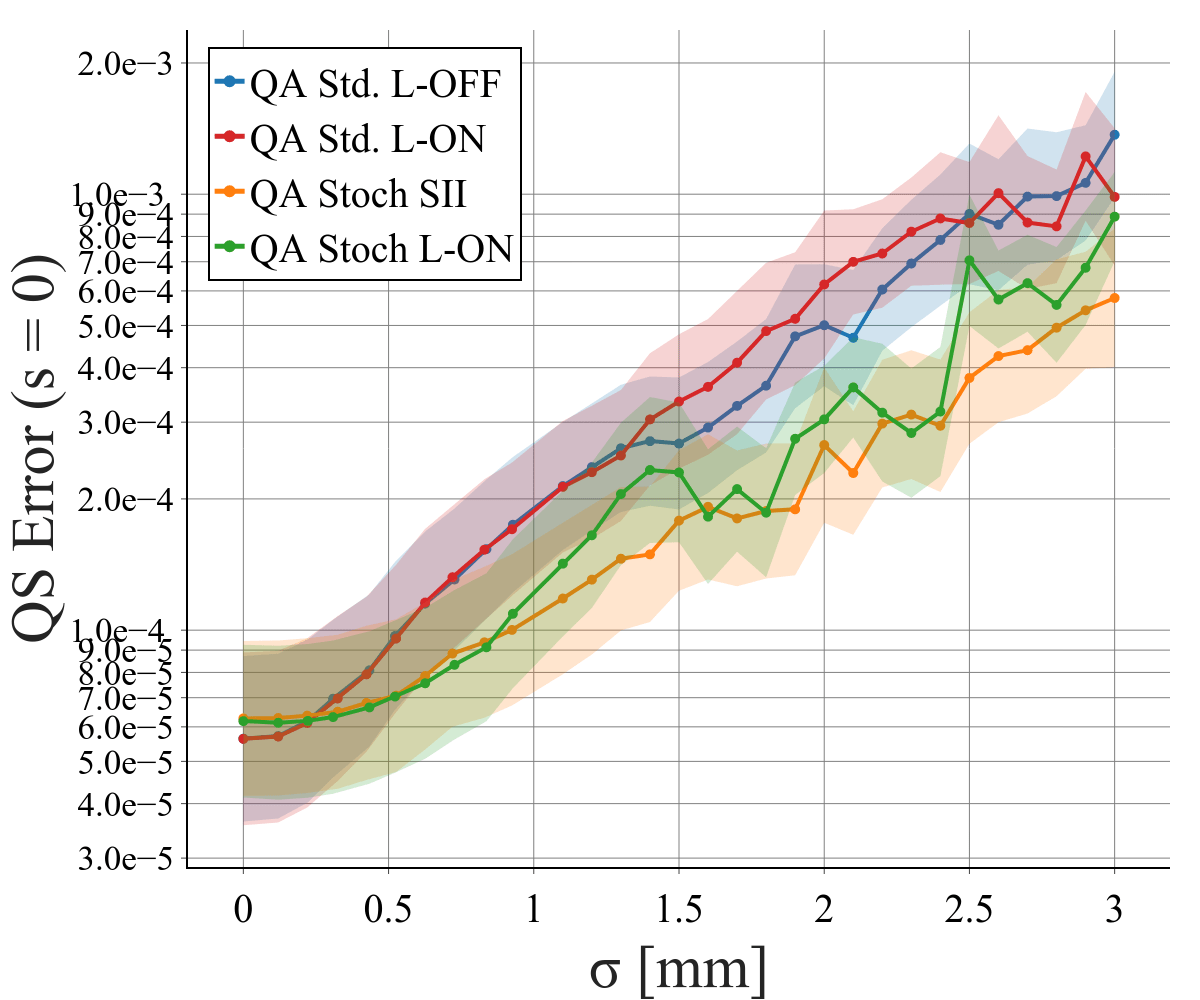}
		\caption{}
		\label{fig:c}
	\end{subfigure}
	\begin{subfigure}[t]{0.4\textwidth}
		\includegraphics[width=\textwidth]{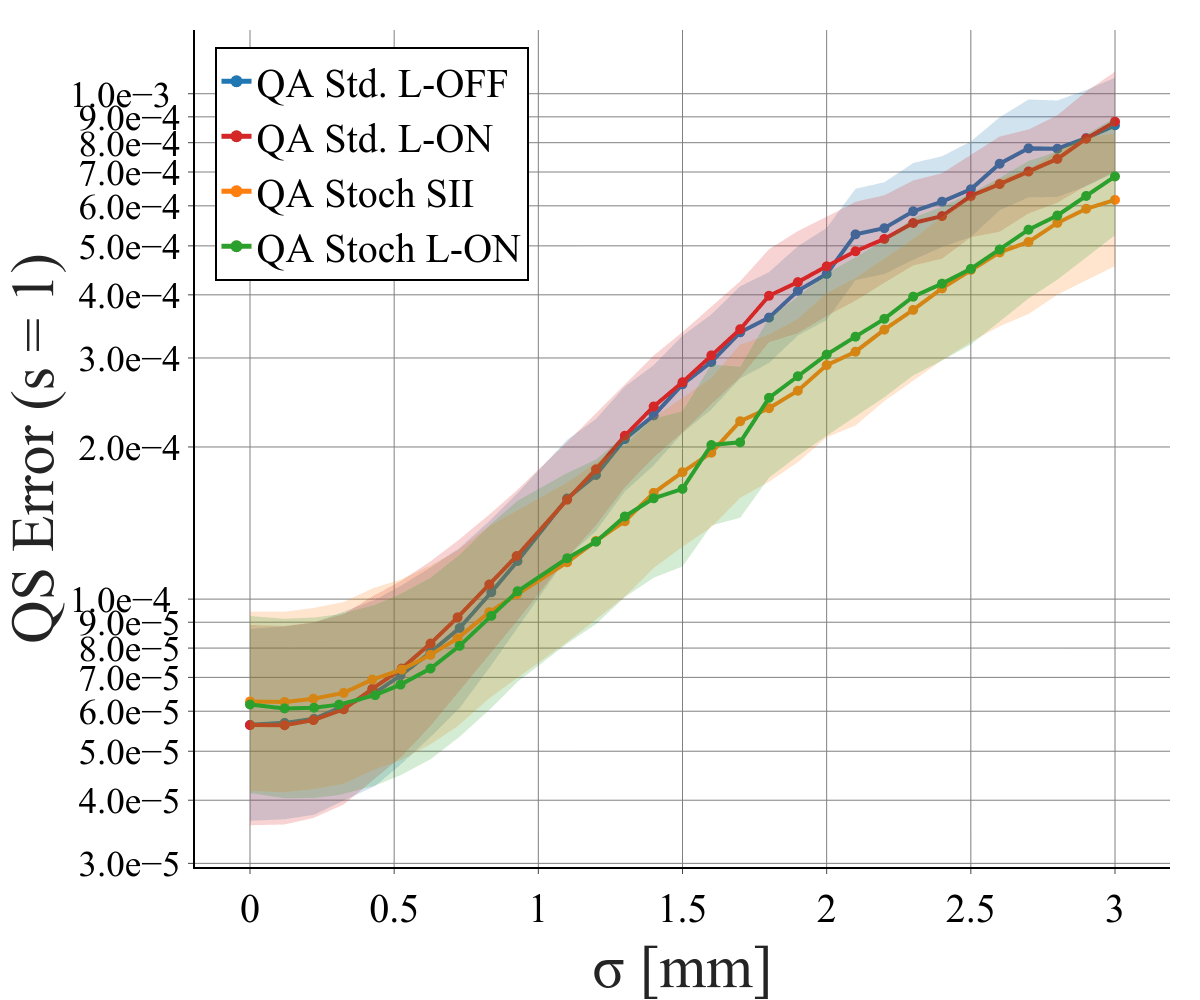}
		\caption{}
		\label{fig:d}
	\end{subfigure}
	
	\caption{Robustness plots for the QA stellarator, both stochastic methods yield comparable results and outperform deterministic approaches. We plot the total (a), core (b), and edge (d) quasisymmetric error versus the coil perturbation $\sigma$ and similarly the squared flux (b). The values of squared flux differ here compared to Table \ref{tab:qa_opts} because here we also normalize with regards to the surface. This also explains why the unperturbed value of the QA Std. configuration appears here with lower SF errors than the stochastic versions.}
    \label{fig:qa_tolerances}
\end{figure}

This is shown in Figure \ref{fig:qa_tolerances} where the total (a) core (c) and edge (d) quasisymmetric errors as well as the squared flux are considered against increasing coil perturbation amplitudes $\sigma$. We see that across all data, both stochastic methods perform equivalently. The robustness profiles of the squared flux are similar to what is displayed in Wechsung et al. \cite{Wechsung_2}, where again the improvements appear small. However, we demonstrate how these can lead to a substantial reduction in the quasisymmetric error for increasing perturbations. While the unperturbed results of the stochastic versions yield slightly worse quasisymmetry than the standard ones, this trend is then reversed at around $\sigma = 0.5$ mm.
Meaning that the standard approach would only be worth employing if the coil manufacturer can ensure a precision lower than around 0.86 mm. This is estimated by multiplying 0.5 mm by $\sqrt{3}$, as $\sigma^2$ only accounts for the pre-exponential factor in one dimension of a 3D space. To the best of our knowledge, until this day, no major stellarator has been built with an accuracy lower than 1 mm. 

\subsection{QA particle confinement}
\begin{figure}[ht]
	\centering
	\includegraphics[width=0.7\linewidth]{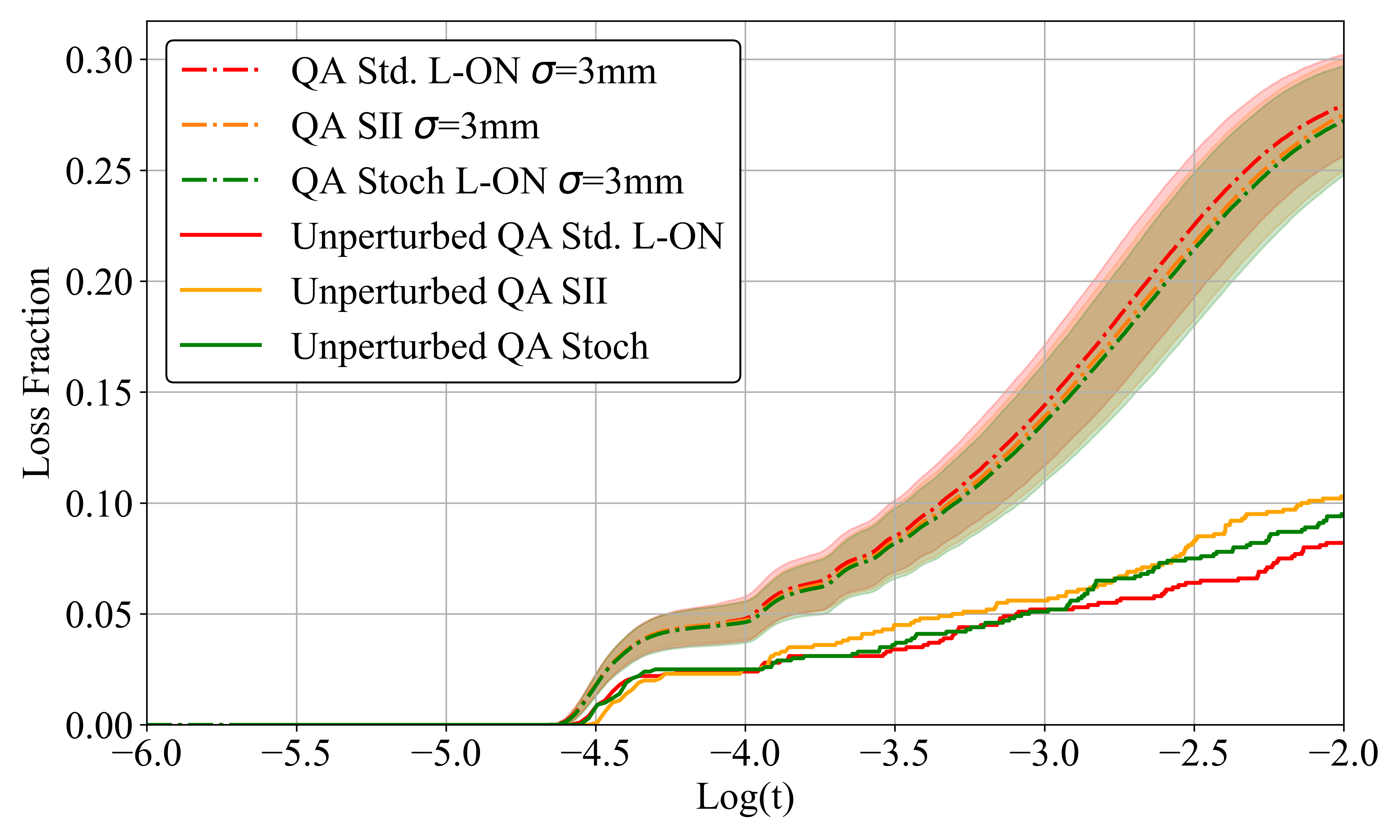}
	\caption{Particle loss fraction comparison of the optimized QA configurations. An average improvement is visible in the loss fraction of both coil sets obtained through stochastic optimization; however, the standard deviations are considerable and dominate the average improvements. Dashed lines correspond to the average of the perturbed configurations and solid lines to the unperturbed stellarators.}
	\label{fig:qa_loss_fraction}
\end{figure}

In this section, we directly inspect the confinement properties of the obtained configurations by performing simulations of particle loss fraction. The goal is to simulate fusion-born $\alpha$-particles in the devices. These are generated with an initial energy of 3.5 MeV, meaning that they can be labeled as collisionless. This means that the quality of quasisymmetry can directly affect their slowing-down time and the amount of energy lost. All the configurations are scaled to be ARIES-CS-like with 5.7 T on axis and equivalent $\rho^{*}=\rho/a$, where $\rho$ is the particle's gyroadius and a its minor radius. Therefore, for the QA case, 1000 particles are initialized with a total energy of 35 keV with uniformly distributed pitch angles, at a radial coordinate of $s=0.3$, and they are followed for 1e-2 seconds. Particles are considered lost if they cross the last closed flux surface at $s=1.0$. The simulations are performed using the SIMPLE code, a guiding-center tracing code that employs a simplectic integrator to estimate particle loss in 3D configurations. In order to quantify $\alpha$ loss robustness, 300 perturbed stellarators at $\sigma$=3 mm are generated and used as the equilibria for the particle loss simulation. These are then compared with the unperturbed equilibria. The results are shown in Figure \ref{fig:qa_loss_fraction}.  

There are two distinct groups of loss fractions. As expected, perturbing the coils by 3 mm and therefore the magnetic field significantly enhances the particle loss (around $\times$2.5 increase). Robustness is assessed as the percentage change in particle loss between the perturbed and unperturbed configurations:

\begin{equation}
	\epsilon_{loss} = \frac{\alpha_{loss}^{pert}-\alpha_{loss}^0}{\alpha_{loss}^0}
	\label{eq:loss_equation}
\end{equation}

In the QA case, we find that $\epsilon_{loss} = 2.4$ for the QA Std. L-ON configuration, and 1.84 and 1.67 for both the QA Stoch and QA SII, respectively. Together with this, and in a similar way as for the QS error robustness plot, the unperturbed version of the standard single stage method outperforms the others for the loss fraction; however, this is reversed at $\sigma=3$ mm.
A key finding from this study is the overall accuracy that can be expected in a built stellarator. As shown in Figure \ref{fig:qa_tolerances}, the unperturbed configurations all reach a squared flux metric of 1e-7 $\text{m}^2$; however, at a mild $\sigma=1.5$ mm deviations, it has deteriorated by two orders of magnitude. The unperturbed accuracy of the stellarator loses its meaning when looking at Figure \ref{fig:qa_loss_fraction}, where slight improvements in field accuracy and even robustness might get blown away by the magnitude of the inaccuracies. This means that stellarator optimization might not need to aim for close to 0 field errors for the \textit{unperturbed} configurations, which comes at the expense of finding increasingly complicated coils, if those incremental improvements are blown away by a mere 1.5 mm perturbations to the coil shapes. Instead, it would be more efficient for new designs to aim for simpler, slightly less accurate coils that exhibit enhanced robustness. 

While in absolute values, the configurations shown here represent considerable losses at a perturbation of 3 mm, the stochastic single-stage approach appears to yield similar results as the previous stochastic stage II method, while also combining the advantages of simultaneous optimization from a single stage. 

\subsection{Quasi-helically symmetric stellarator}

In this section, we present the results from a stochastic single-stage optimization made with a QH configuration. Similarly to the previous section, all the single-stage optimizations presented in this work are performed following a warm-start stage II optimization to facilitate the simultaneous minimization of both the surfaces and the coils. Here, we make three optimizations instead of four: a standard single-stage run (QH Std.), a stochastic stage II run (QH Stoch SII) with the warm-start surface, and a stochastic single-stage run (QH Stoch). For all the runs here, we perform the optimization including all the constraints defined in Equation \ref{eq:coil_functions}. A QH configuration is, by default, more demanding from a coil perspective due to its complex shape. We choose to optimize here for a four-field-period QH configuration with a 1 m major radius, aspect ratio of 7, and iota of 1.19. The target values kept throughout the optimization are 7 for the aspect ratio and 1 for the rotational transform. For the single-stage runs, the threshold for coil-to-surface distance is set at 7 cm, coil-to-coil distance at 6 cm, maximum curvature at 15 $\text{m}^{-1}$ and maximum mean-squared-curvature (MSC) at 5 $\text{m}^{-1}$. For the stage II coil set, MSC and maximum curvatures are both relaxed to 20 $\text{m}^{-1}$ as it proves difficult to find configurations without interlocked coils. Note that, in this study, we allow for a considerable amount of constraint violation of the regularization functions, as the main goal is to show how robustness can be improved. This assumes that robustness is independent of curvature, coil-to-coil distance (with cc-distance $\gg \sigma$), and other configuration features, except for coil length. Indeed, the QH Stoch SII configuration is designed to match the specifications of the QH Stoch coil set, making the robustness comparison between the two valid. 

\begin{table}[ht]
	\centering
	\begin{tabular}{|c|c|}
		\hline
		Coil and Eq. Properties                     & Warm-start QH \\ \hline
		$\langle B\cdot n\rangle / \langle B \rangle$ &   3.79e-3                                \\ \hline
		QS error (Fixed-Boundary) [a.u]                             & 3.66e-4                              \\ \hline
		Total Coil Length [m]                               & 12.5                                     \\ \hline
		CS distance [m]                       &  0.10     \\ \hline
		CC distance [m]                       &    0.062 \\ \hline
		Max Curvature $\kappa$ [$\text{m}^-1$] &    9.76      \\  \hline
		Max MSC Curvature [$\text{m}^-1$] &     15.00  \\ \hline
		Mean $\iota$                        &  1.19                                            \\ \hline
		Aspect Ratio                        &  7                                            \\ \hline
	\end{tabular}
	\caption{Coil and equilibrium parameters of the warm-start configuration used for the QH stellarator. Here, the total coil length designates the total coil length per half-field period.}
	\label{tab:warm_start_properties_qh}
\end{table}

As shown in Table \ref{tab:warm_start_properties_qh}, the initialized coils already show a relatively low $\langle B\cdot n\rangle / \langle B \rangle$, similarly for the QS error of the equilibrium in fixed-boundary mode. The primary requirement for the initialization coil set is that the squared flux is sufficiently low to prevent poor optimization and that there are no interlocked coils. There is still some freedom that is allowed in all the remaining constraints to try and exploit the single-stage link between equilibrium and coils during the optimization. Whereas in the QA case, the minimization of the departure from quasisymmetry of the equilibrium is desired, here we opt for a different approach that is mainly driven by the squared flux. This means that if the resulting quasisymmetry from the single-stage step is worse than the warm-start solution, it is still retained. The motivation behind this is similar to the argument given previously, where it is explained that perturbations will, in any case, deteriorate any slight improvements in quasisymmetry. 

\begin{figure}[ht]
	\centering
	
	
	\begin{subfigure}[t]{0.4\textwidth}
		\includegraphics[width=\textwidth]{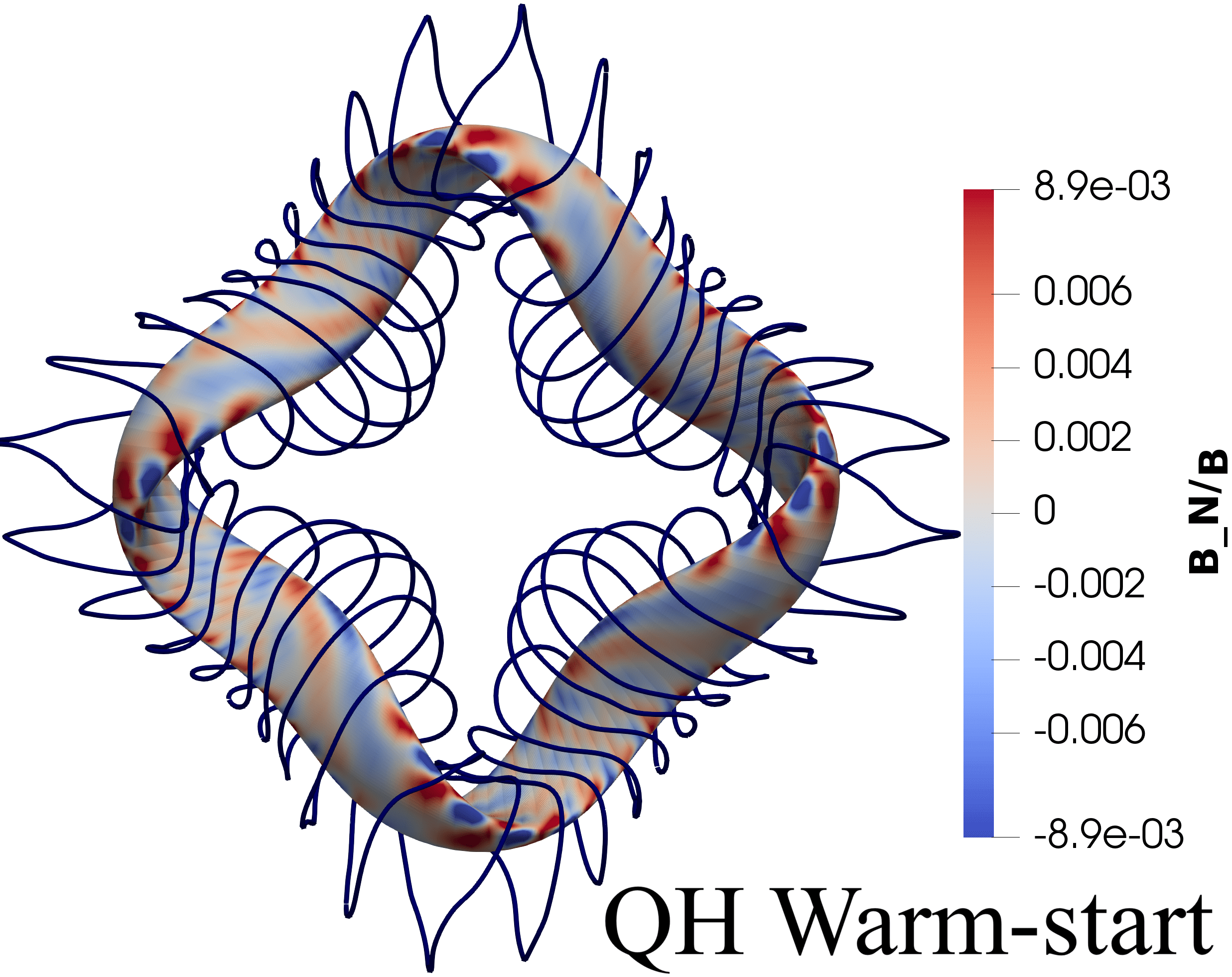}
		\caption{}
		\label{fig:coils_warm_start}
	\end{subfigure}
	\hspace{1cm}
	\begin{subfigure}[t]{0.4\textwidth}
		\includegraphics[width=\textwidth]{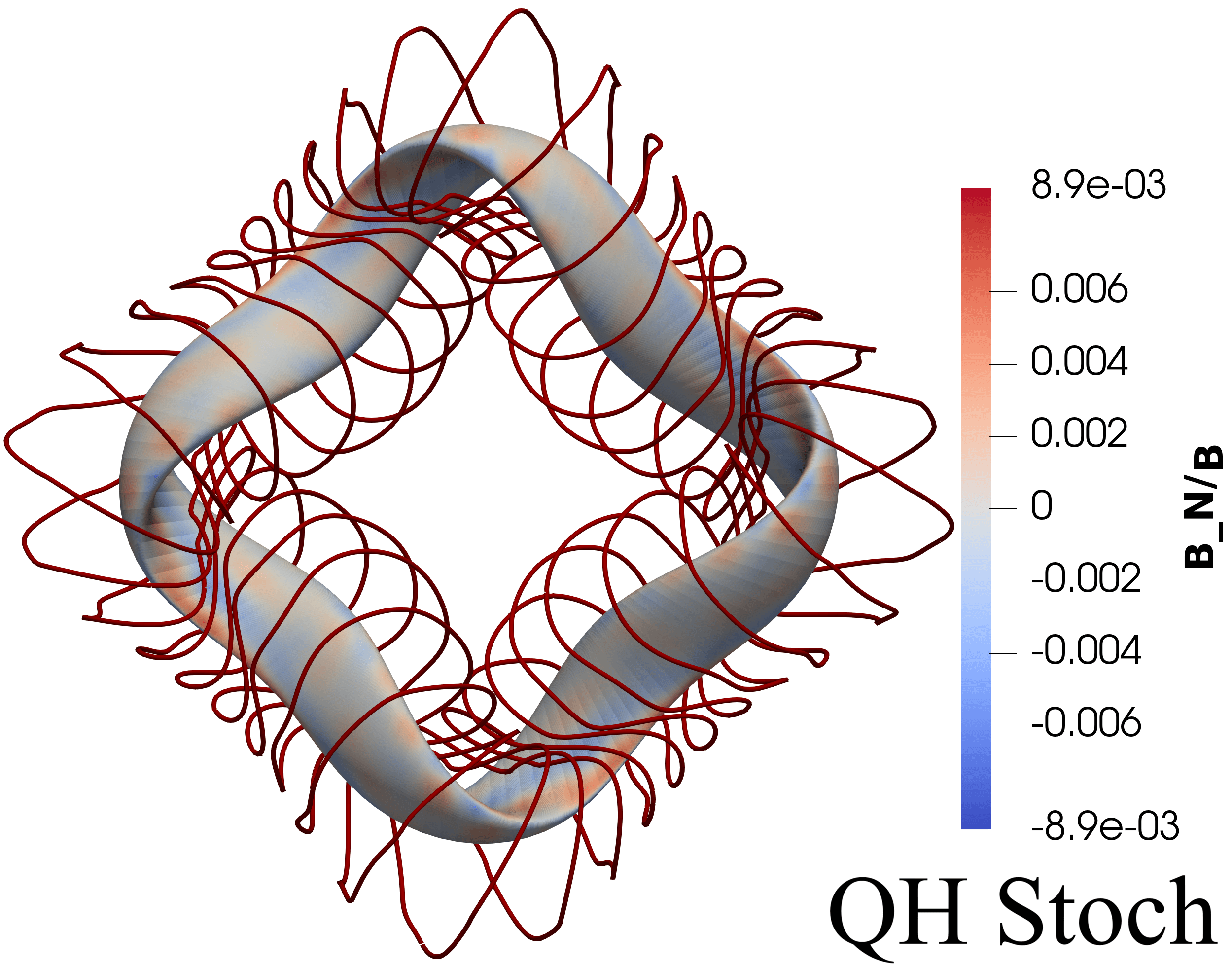}
		\caption{}
		\label{fig:coils_stoch}
	\end{subfigure}
	
	\par\bigskip
	
	\begin{subfigure}[t]{0.4\textwidth}
		\includegraphics[width=\textwidth]{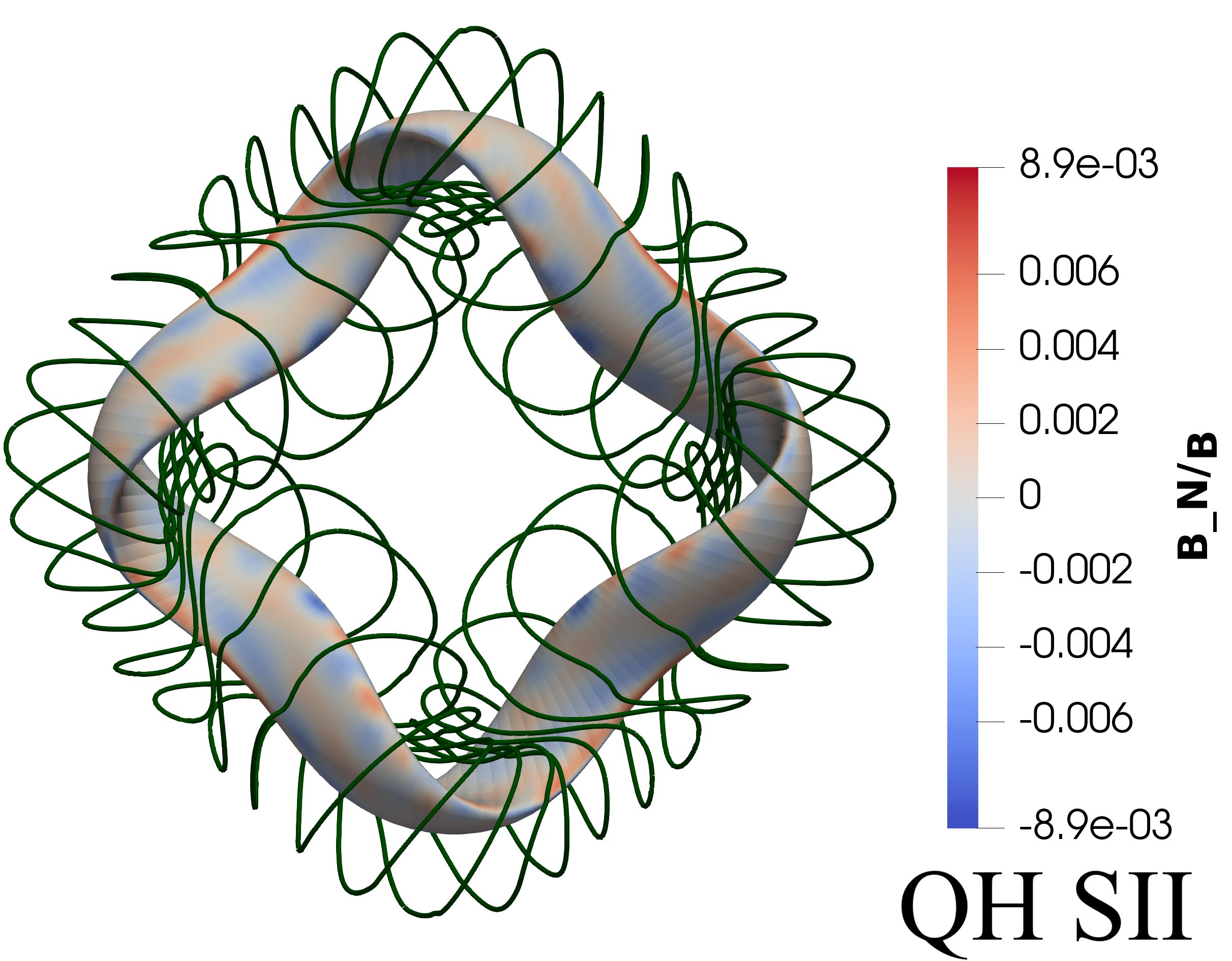}
		\caption{}
		\label{fig:coils_SII}
	\end{subfigure}
	\hspace{1cm}
	\begin{subfigure}[t]{0.4\textwidth}
		\includegraphics[width=\textwidth]{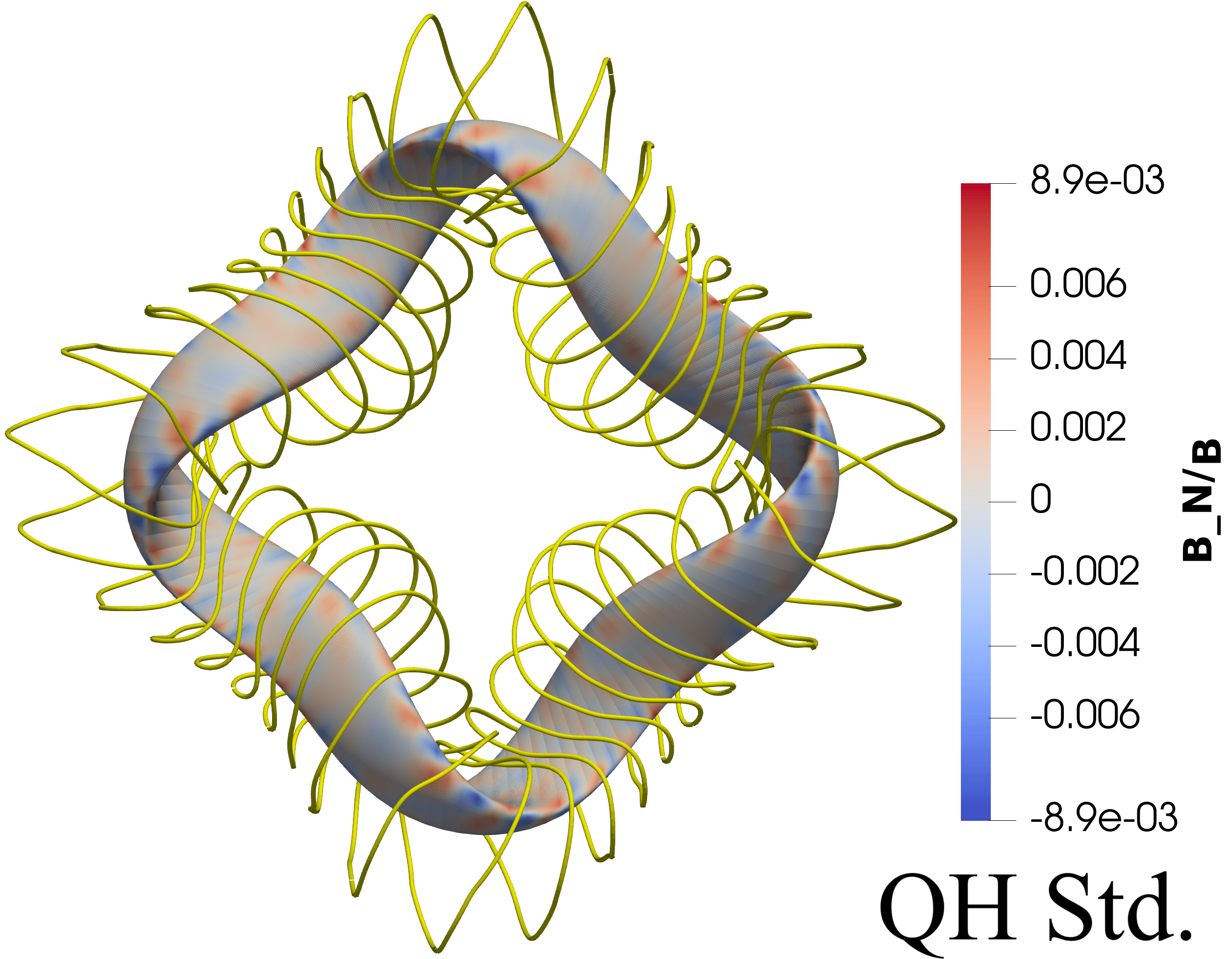}
		\caption{}
		\label{fig:d_qs_qh}
	\end{subfigure}
	
	\caption{ Results of the stochastic single stage method applied to the quasi-helically symmetric case. Coils obtained from a stage II warm-start (a), coils obtained from stochastic single stage (b),  coils obtained from stochastic stage II (c), coils obtained from standard single stage (d).}
	\label{fig:qh_coils}
\end{figure}

For the stochastic single-stage run, increasing $\sigma$ to 5 mm, instead of 2.5 mm for the QA, yields optimal results; the length L is kept the same. The stellarators resulting from the subsequent optimizations are shown in Figure \ref{fig:qh_coils}, including the warm-start coils. The properties of the corresponding optimized configurations are shown in Table \ref{tab:qh_opts}. Again, here the QH Std. The set remains closely resembling the starting coil set, with only slightly longer coils; however, the squared flux is improved. This is likely due to a combination of relaxation of the coil geometric constraints, such as curvature and critical distances, as well as a substantial increase in the quasisymmetric error ($\times$2.5). For both stochastic runs, lower levels of field error are achieved, along with lower quasisymmetric errors, as indicated by the QH Std. run. This applies to both QS errors in fixed-boundary and from the QFM surfaces. However, they exhibit higher constraint violations for curvature and coil-to-coil distances. This results in sections with increased coil complexity around the corners of the stellarator as shown in Figure \ref{fig:qh_coils}b) and c).

\begin{table}[ht]
\centering
\begin{tabular}{|c|c|c|c|}
\hline
    & QH Std. & QH Stoch SII & QH Stoch \\ \hline
Max $(B\cdot n) /  B $ &                                           8.87e-3               &         6.53e-3           &      \textbf{5.89e-3}      \\ \hline
$\langle B\cdot n\rangle / \langle B \rangle$ &           1.68e-3                                  &                   1.49e-3         &      \textbf{1.32e-3}      \\ \hline
Total Coil Length [m]                    &      \textbf{12.63}          &       14.19          &    14.27       \\ \hline
CS distance [m]                       &  0.12     &    \textbf{0.13}       &     \textbf{0.13}                \\ \hline
CC distance [m]                       &    \textbf{0.058}        &      0.038         &  0.056             \\ \hline
Squared Flux Metric [m$^{2}$]                   &    1.55e-5       &   9.95e-6      &      \textbf{8.60e-6}          \\ \hline
Max Curvature $\kappa$ [$\text{m}^-1$] &    \textbf{12.52}         &   59.83        &  18.72     \\  \hline
Max MSC Curvature [$\text{m}^-1$] &     \textbf{16.73}    &     23.00       &   18.3   \\ \hline
QS error (Fixed-Boundary) [a.u]               &     8.64e-4      &   3.66e-4*     &      \textbf{3.94e-4}          \\ \hline 
QS error (QFM) [a.u]                   &    \textbf{2.7e-3}    &    4.4e-3      &      2.9e-3          \\ \hline 
QS at $\sigma =$ 1.5 mm   [a.u.]           &    4.1e-2        &      1.4e-2            &   \textbf{1.3e-2}          \\ \hline 
Average iota & 1.19 & 1.19* & 1.2 \\ \hline 
Aspect Ratio & 7.0 & 7.0* & 7.1 \\ \hline 
\end{tabular}
\caption{Properties of both QH coils and the equilibria that are given to the optimizer for the QH stellarator in the three different optimizations. Here we compare the stochastic single-stage (Stoch QH) to both traditional stage I and stochastic stage II optimization (Stoch QH-SII) and standard single-stage (Std. QH) according to \cite{Jorge_2023}. QS at $\sigma=3$ mm refers to an average quasisymmetric error measured for samples perturbed with a coil deviation amplitude of about 3 mm. 
*These values correspond to the values of the warm-start stage I equilibrium shown in Table \ref{tab:warm_start_properties_qh}.}
 \label{tab:qh_opts}
\end{table}

\begin{figure}[ht]
	\centering
	\includegraphics[width=0.47\linewidth]{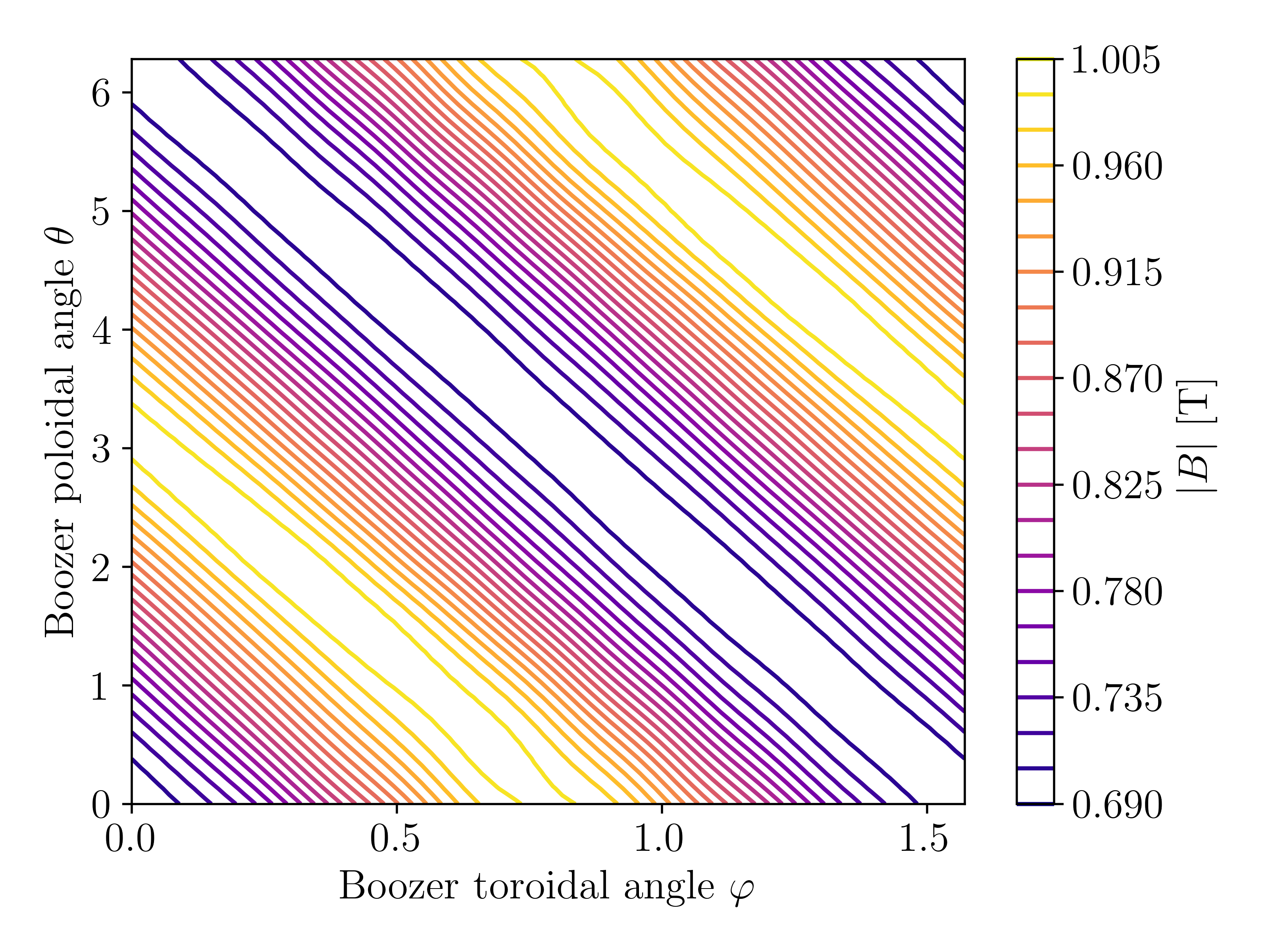}
	\includegraphics[width=0.47\linewidth]{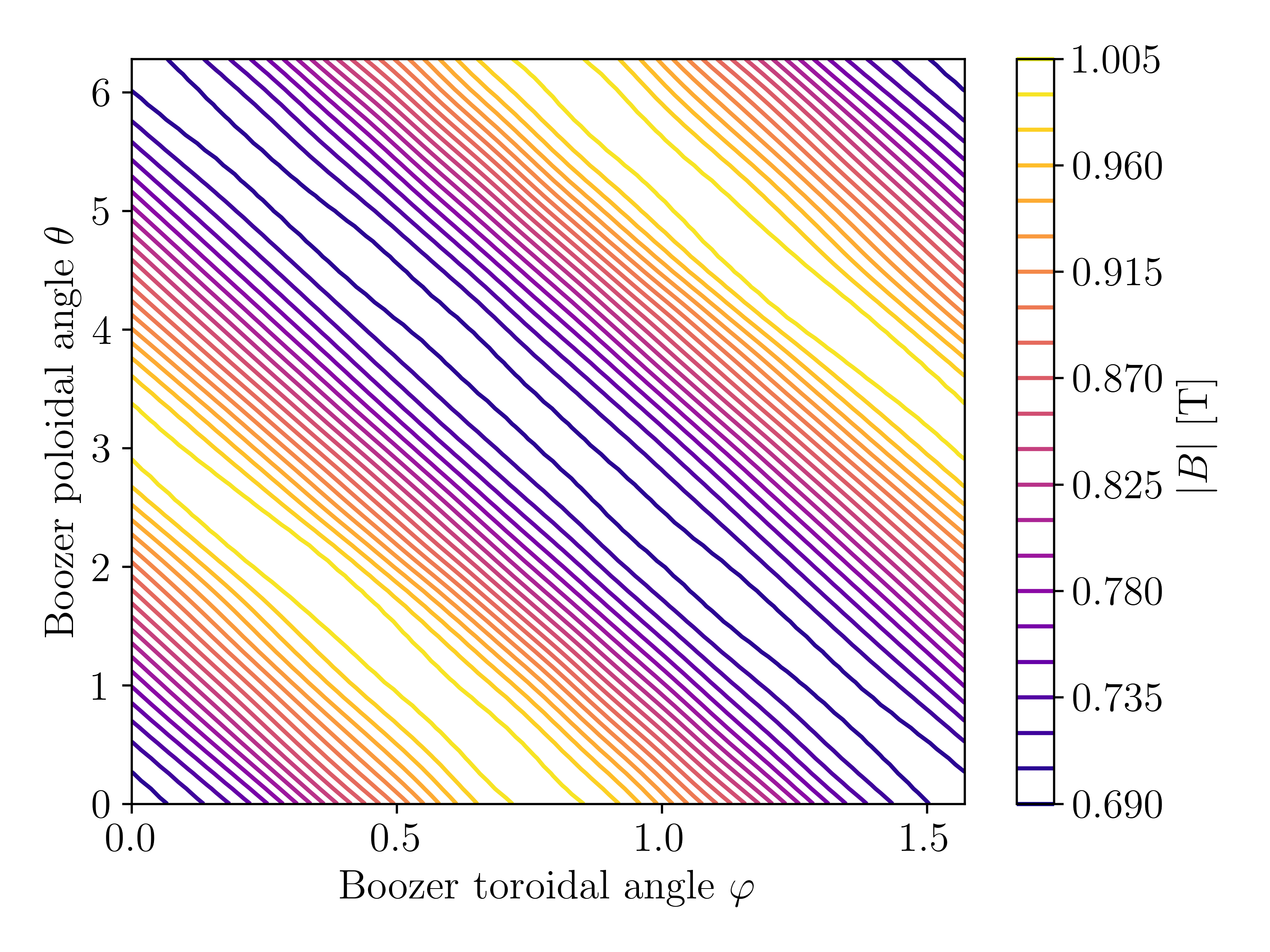}
	\caption{\textbf{Left:} Boozer plot at s=0.9 displaying qualitatively the departure from quasisymmetry from the unperturbed stochastic single stage QH equilibrium calculated using QFM surfaces. No coil ripple is visible, and mostly straight contours are visible, showcasing an example of good quasisymmetry. \textbf{Right:} Boozer plot at s=0.9 of the same coil set but perturbed with $\sigma = 3$ mm.}
	\label{fig:qh_boozer}
\end{figure}

\begin{figure}[ht!]
    \centering
    \includegraphics[width=0.45\linewidth]{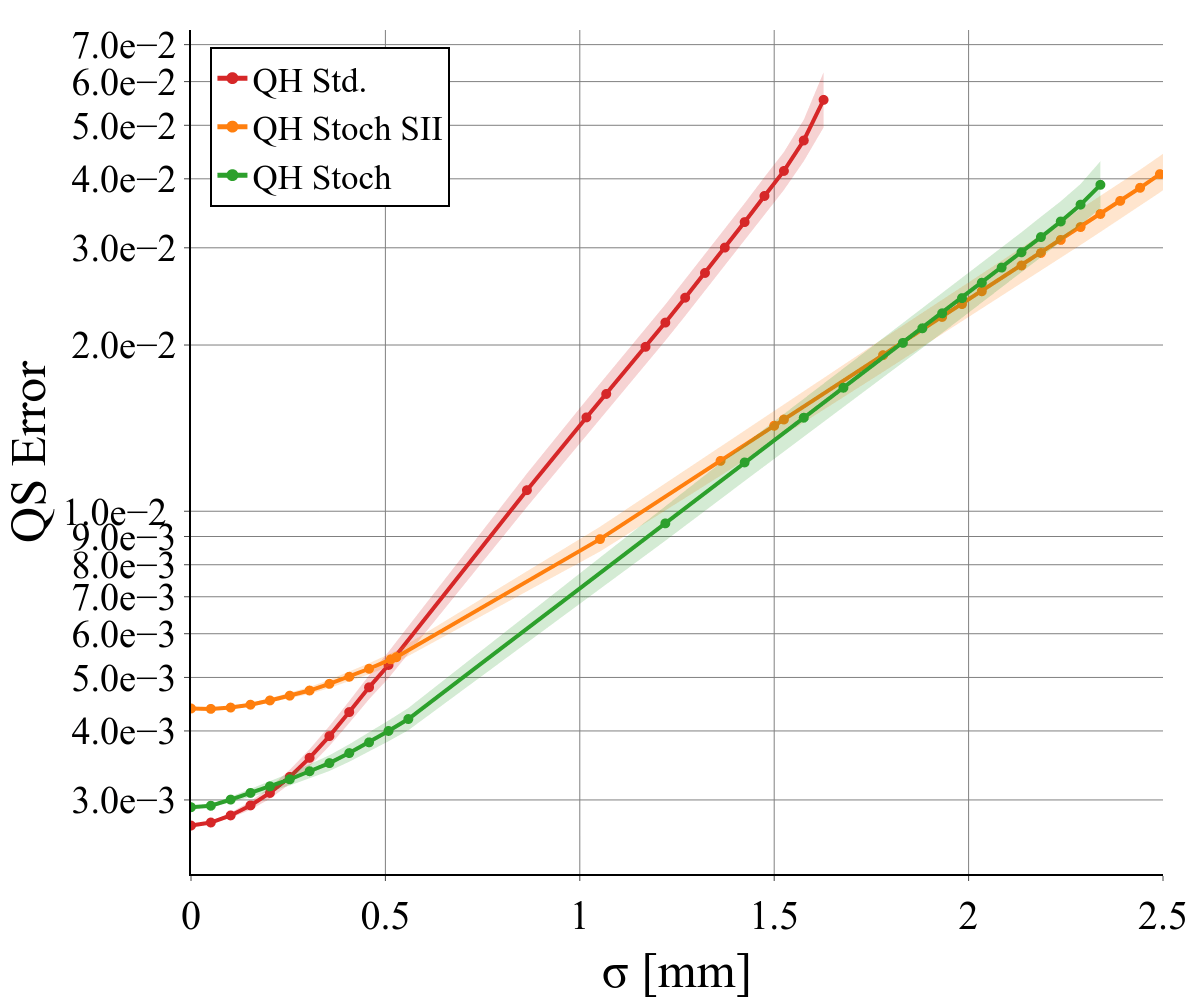}
    \includegraphics[width=0.45\linewidth]{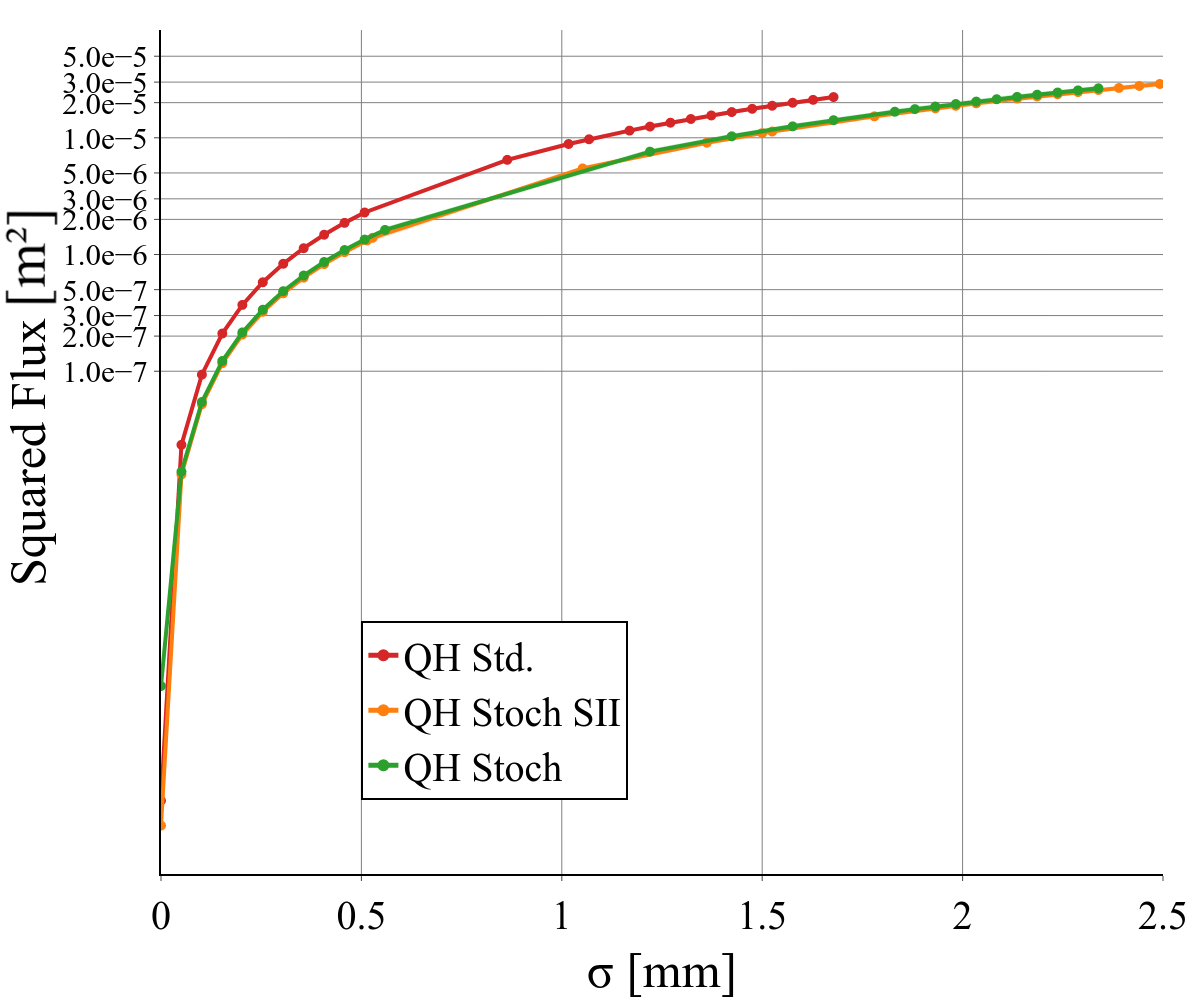}
    \caption{ Tolerance plots for the QH stellarator, both stochastic methods yield comparable results and outperform deterministic approaches. We plot the total quasisymmetric error (\textbf{left}) and squared flux (\textbf{right}) versus the coil perturbation $\sigma$. Here, the unperturbed value of the QH Std. configuration appears with lower SF errors than the stochastic versions (unlike Table \ref{tab:qh_opts}) due to a different surface normalization, which yields variations of the order of 1e-8.}
    \label{fig:qh_tolerances}
\end{figure}

 Nevertheless, this results in improved robustness considering perturbations at $\sigma=1.5$ mm, where both stochasticly optimized configurations show quasisymmetric errors that are about 3 times lower at around 1\%. In Figure \ref{fig:qh_boozer}, the Boozer plots at the LCFS of the \textit{unperturbed} and \textit{perturbed} coil set obtained from stochastic single-stage are shown. Little coil ripple is noticeable, and straight diagonal lines are still present, meaning that quasisymmetry is conserved to a high degree even at a perturbation of $\sigma=1.5$ mm. This far improves on the QA case, where at 3 mm perturbation there is only a slight improvement of the quasisymmetric error $\sim$0.3\% compared to the deterministic method. The respective robustness plots are shown in Figure \ref{fig:qh_tolerances}. We observe a similar trend as shown for the QA case in Figure \ref{fig:qa_tolerances}, where the standard method performs marginally better for the unperturbed state than both stochastic configurations. This order, however, is inverted in the quasisymmetric error for the stochastic single-stage configuration at around 0.25 mm and then later for the stochastic stage II coil set at 0.5 mm. The influence of improved robustness appears even earlier for the squared flux. Both stochastic single-stage and stochastic stage II exhibit similar slopes in their squared flux and quasisymmetric error; however, at about 2.25 mm, the QH Stoch II case shows marginally lower quasisymmetric errors. 

 \subsection{QH particle confinement}

\begin{figure}[ht]
 	\centering
 	\includegraphics[width=0.7\linewidth]{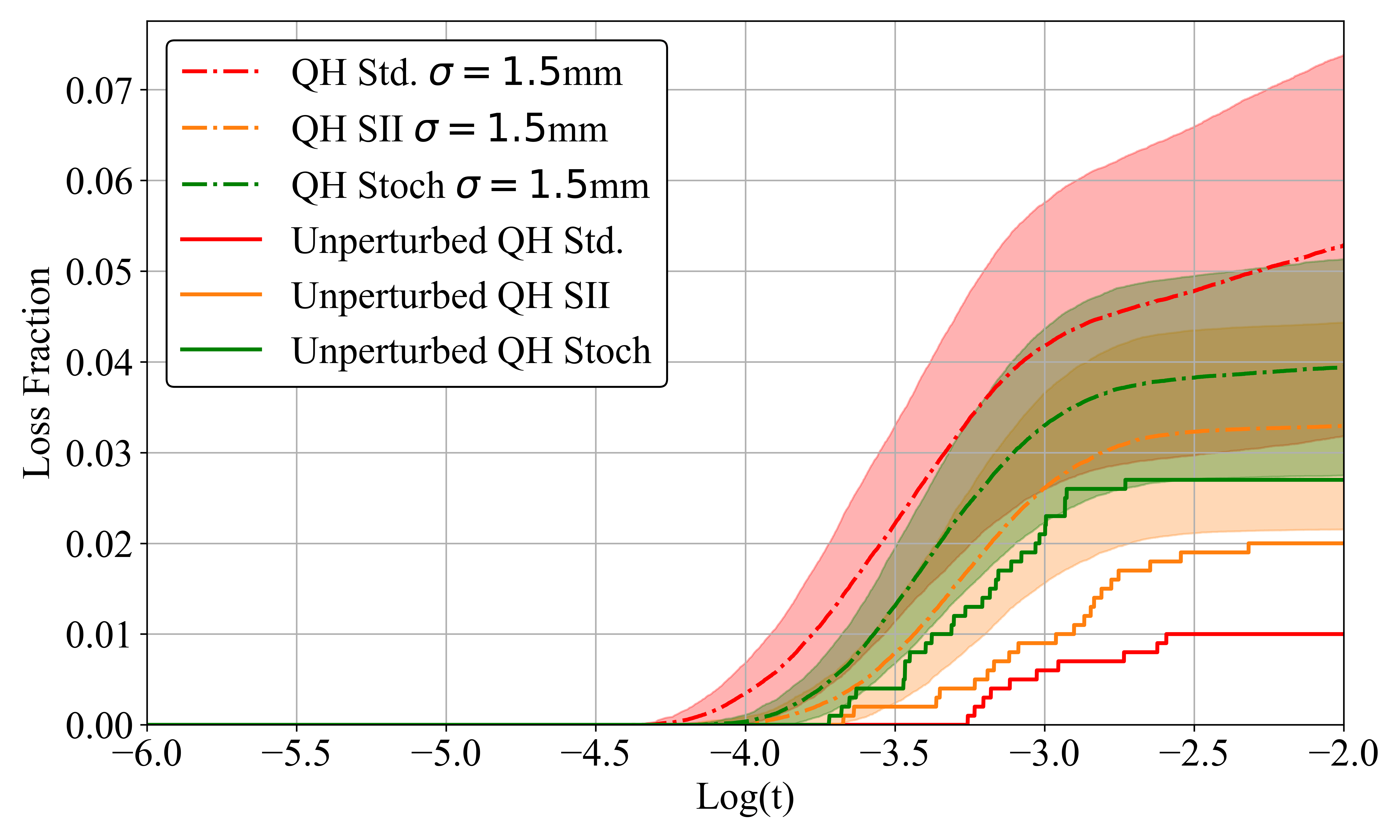}
 	\caption{Particle loss fraction comparison of the optimized QH configurations. Here, both average and standard deviations are improved for the stochastic methods; moreover, the stochastic single-stage method (here QH Stoch) shows better overall robustness.}
 	\label{fig:qh_loss_fraction}
 \end{figure}

 In a similar manner to the previous QA case, we estimate the particle loss here by scaling the simulation parameters to achieve a similar $\rho^{*}$ to that of the ARIES-CS reactor. Given that this configuration possesses an average minor radius of 0.14 m, the energies are reduced to 25 keV. Moreover, for the QH case, we start the particles further out radially (at $s=0.6$) than previously. This is done to observe particle losses after 1e-2 s, given that the confinement of trapped particles in a high-iota QH configuration is improved when compared with the QA case at similar QS errors. The results are plotted in Figure \ref{fig:qh_loss_fraction}.
 
 Unlike the QA stellarators, where the gains in quasisymmetry are marginal, here there is a visible improvement in particle loss. This is shown by the average of the 300 perturbed stellarators that reaches 5.3\% at $\sigma=1.5$ mm for the QH Std. configuration, whereas the stochastic version gives 3.9\% and 3.3\% for the QH SII and QH Stoch, respectively. This aligns with the previously observed trend, where the QH Std. version has \textit{lower} loss fraction values than the other configurations for $\sigma=0$ mm, however this reverses already at $\sigma=1.5$ mm. Moreover, if we calculate the proportional difference of the loss fraction as described in Equation \ref{eq:loss_equation}, we find that the QH Std. has an increase of particle loss of about 430\%, whereas the QH SII has a 65\% increase and the QH Stoch coil set ends up with the lowest increase of particle loss at 44\%, which is one order of magnitude better than the standard case. 
 
\section{Conclusion}
\label{sec:conclusion}

In this work, we have presented the results of merging two existing methods: stochastic coil optimization and single-stage optimization. Results show that it is possible to combine the advantages of each method into a single one, namely the increased robustness from the stochastic method, together with simultaneous equilibrium and coil optimization from a single stage, methods, for enhanced compatibility of the coils with the plasma surface. First, it is demonstrated that increasing the perturbation amplitude, which corresponds to an increase in the squared flux cost function, does not impact the performance of the optimization of other cost functions. This is traditionally not the case when increasing the amplitude of the squared flux hyperparameter. 
Then, the method is benchmarked against standard deterministic single-stage and stochastic coil optimization methods in two different quasisymmetry configurations: one quasiaxisymmetric, one quasi-helically symmetric. These are performed starting from already optimized stellarators to guide the optimization and investigate whether improvements are possible. Both configurations demonstrate that the stochastic single-stage method achieves comparable quasisymmetry and field error robustness to the stochastic stage II method, while consistently attaining better equilibrium quasisymmetry and squared flux than its standard single-stage counterpart. This is explained by the standard method getting quickly stuck in the local minimum provided by the 'warm-start'. In contrast, the stochastic single-stage approach manages to escape the local minimum and explore the optimization space further. 
Finally, while the gain in robustness is marginal in the QA case, $\alpha$-particle loss simulations confirm the observed trend of improved robustness, even if for similarly marginal gains at 3 mm perturbations. Note that the marginal improvements are also observed for the stochastic stage II configuration. Nevertheless, a threefold QS error improvement is observed over the standard single-stage method for the QH case. This is additionally demonstrated by similar $\alpha$-particle loss simulations yielding a 44\% increase in loss for the stochastic single-stage QH coil set versus a 430\% increase for the standard method. 
We can therefore conclude that this method effectively combines the advantages of both approaches and can serve as a functional stellarator optimization approach. The compatibility between the equilibrium and the coil set still could be improved by transitioning from a fixed-boundary equilibrium optimization to a free-boundary approach, where it is possible to differentiate equilibrium metrics with regard to coil degrees of freedom. This study also demonstrates that stellarator coil optimization does not require aiming for asymptotically minor field errors, as real-world perturbations to the coils will blow away any marginal improvements in accuracy.

%
%

\ack{We thank Alan Kaptanoglu, Florian Wechsung and Matt Landreman for useful discussions.}

\funding{This work was supported through grants from the Helmholtz Association Young Investigators Group program as project VH-NG-1430.}

\data{Sample text inserted for demonstration.}

\bibliographystyle{ieeetr}
\bibliography{bibliography}

\end{document}